\documentclass[pre,reprint]{revtex4-1}  % doubl-spacing

\usepackage{graphicx}  % needed for figures
\usepackage{dcolumn}   % needed for some tables
\usepackage{bm}        % for math
\usepackage{amssymb}   % for math

\begin{document}

\title{Effective magnetic susceptibility of suspensions}

\author{Kunlun Bai}
\affiliation{Department of Mechanical Engineering and Materials Science, Yale University, New Haven, CT 06511, USA}

\author{Joshua Casara}
\affiliation{School of Natural Sciences, University of California, Merced, CA 95343, USA}

\author{Aparna Nair-Kanneganti}
\affiliation{Department of Mechanical Engineering and Materials Science, Yale University, New Haven, CT 06511, USA}

\author{Aubrey Wahl}
\affiliation{Department of Mechanical Engineering and Materials Science, Yale University, New Haven, CT 06511, USA}

\author{Florian Carle}
\affiliation{Department of Mechanical Engineering and Materials Science, Yale University, New Haven, CT 06511, USA}

\author{Eric Brown}
\email{eric.brown@yale.edu}
\affiliation{Department of Mechanical Engineering and Materials Science, Yale University, New Haven, CT 06511, USA}
\affiliation{School of Natural Sciences, University of California, Merced, CA 95343, USA}

\date{\today}%Please keep this in while we are working on drafts -  it will automatically keep ordering

\begin{abstract}
We characterize how suspensions  of magnetic particles in a liquid  respond to a magnetic field in terms of the effective magnetic susceptibility $\chi_{eff}$ using inductance measurements.  We test a model that predicts how $\chi_{eff}$ varies due to demagnetization, as a function of sample aspect ratio, particle packing fraction, and particle aspect ratio \cite{skomski_effective_2007}.  For spherical particles or cylindrical particles aligned with external magnetic field, the model can be fitted to the measured data with agreement within 17\%. However, we find that the random alignment of particles relative to the magnetic field plays a role, reducing $\chi_{eff}$ by a factor of 3 in some cases, which is not accounted for in models yet.  While suspensions are predicted to have  $\chi_{eff}$ that  approach the particle material  susceptibility in the limit of  large  particle aspect ratio,  instead we find a  much smaller particle aspect ratio where  $\chi_{eff}$  is maximized.  A  prediction that  $\chi_{eff}$ approaches the bulk material  susceptibility  in the limit of the packing fraction of the liquid-solid transition also fails.  We find $\chi_{eff}$  no larger than about 4 for suspensions of iron particles.  
\end{abstract}

\maketitle

\section{Introduction}

 %motivation
 Suspensions  of magnetic particles in a liquid can be controlled by an applied magnetic field,  a property that is taken advantage of for example in  the fields of ferrohydrodynamics \cite{rosensweig_ferrohydrodynamics_2013}  and magnetorheology \cite{vicente_magnetorheological_2011}. The parameter that directly controls the force applied by a magnetic field in these cases is the effective magnetic susceptibility $\chi_{eff}$. If the suspensions are also conducting, magnetohydrodynamic effects can occur, such that a magnetic field can in principle be generated by  the conducting fluid flow, and the magnetic field can deflect the conducting flow via a Lorentz force, effects whose magnitude scales with $1+\chi_{eff}$ \cite{monchaux07,stieglitz01,gailitis00}.  While these  phenomena  are not easily achieved with known fluids -- where pure conducting liquids generally have a magnetic  susceptibility $\chi\ll 1$ --  there is potential that if a material can be designed with large enough $\chi_{eff} \stackrel{>}{_\sim} 1$, these phenomena could be more easily observed on a laboratory or device scale of order 10 cm \cite{CBCVB17}. Our goal is to determine how the effective magnetic susceptibility $\chi_{eff}$ depends on the particle properties of suspensions.  In particular, we would like to obtain larger values of $\chi_{eff}$ to make such suspensions useful for producing magnetohydrodynamic phenomena on the laboratory scale.   

The effective susceptibility $\chi_{eff}$ is defined by the proportionality $\chi_{eff} = \phi M/H_{app}$, where $H_{app}$ is an externally applied magnetic field, $M$ is  the magnetization per unit volume of magnetic material, and $\phi$  is the volume fraction of the magnetic particles. Note that the factor of $\phi$ in the expression  differs from traditional definitions in pure materials where susceptibility is  defined per unit volume of magnetic material, as a pure material is 100\% magnetic material. Instead we define $\chi_{eff}$ as susceptibility per unit volume of sample, since we are interested in the force from an applied  magnetic field on the sample as a whole.   For linear magnetic materials $\chi_{eff}$ is independent of $H_{app}$, in practice this tends to be the case for small $H_{app}$ before the magnetization begins to saturate.

Locally,  the magnetic susceptibility $\chi=M/H$  is considered a bulk material property depending on the local magnetic field $H$.  In contrast, $\chi_{eff}$  as a macroscopic parameter can be much smaller than $\chi$ due to demagnetization,   an effect in which the  induced magnetic dipole creates  an additional magnetic field $DM$ (where $D$  is called the demagnetization factor) that opposes $H_{app}$.  The net magnetic field inside the material $H_{app}-DM$ that determines the net local magnetization $M$ is less than $H_{app}$,  resulting in $\chi_{eff}$ being smaller than $\chi$.   The effective susceptibility can then be written  such that the demagnetization is a correction factor on the material susceptibility:
\begin{equation}    
\frac{\phi}{\chi_{eff}} = \frac{1}{\chi} + D \ . 
\label{Eq_sus_eff_mat_demag}
\end{equation}

% demagnetization for single piece solid magnets
  It is well-known for single-piece solid magnets, for example,  that $D$  depends on the shape of the magnet,  in particular $D$ is small  in the limit of long,  thin magnets aligned with the applied magnetic field,  (in  this limit $\chi_{eff}$ approaches the material susceptibility $\chi$).   For single-piece solids, unless the aspect ratio of the material is extremeley large, $\chi_{eff} \ll \chi$ and  to a good approximation $\chi_{eff} \approx \phi/D$.  $\chi_{eff}$ has been calculated for many particle shapes \cite{aharoni_demagnetizing_1998}.    For example, for a spherical particle  $D=1/3$, resulting in a maximum $\chi_{eff} \approx  3$ as long as $\chi\gg 3$.  For such geometries with aspect ratio close to 1, the demagnetization effect can be considered a dominating factor determining $\chi_{eff}$, rather than a small perturbation on the material susceptibility $\chi$.
  
  %and $\chi_{eff} \approx  3\phi$ for non-interacting particles in a suspension  as long as $\chi\gg 3$. %( cite textbook?, \textbf{i need to find this}).  
%which is analogous to the approach used by Landau and Lifshitz to estimate the electric permittivity of a suspension of spherical particles \cite{frick_effective_2002}. The Landau-Lifshitz relation only works for very small packing fraction but it cannot describe the behavior at high packing fraction. 

%previous work-suspensions
Demagnetization factors are less well-understood for systems of random arrangements of particles such as suspensions.  With many particles, the demagnetization factor $D$ can depend on  geometries of both the particles and the sample as a whole, as well as positions and alignments of particles relative to each other and the applied magnetic field.

%MR
For example, in the magnetorhoelogical effect, a suspension exposed to an applied magnetic field develops a yield stress.  The magnitude of this yield stress scales roughly as the force of  the induced dipole-dipole interaction between the particles in suspension, which is proportional to $\chi_{eff}^2$  \cite{vicente_magnetorheological_2011}.   It has been observed that large-aspect ratio rod-shaped particles exhibit a larger yield stress than spherical particles, which was qualitatively attributed to the demagnetization effect \cite{vicente_effect_2010}.   However,  this is not quantitatively understood, due to the lack of a model that relates this yield stress to $\chi_{eff}$ and the  demagnetization effect.

%phi, gammas
For randomly packed spherical particles, it has been theoretically argued that the demagnetization factor is $D = \frac{1}{3}+\phi(D_g-\frac{1}{3})$, where $D_g$ is the global demagnetization factor based on the geometry of the sample and $\phi$ is the volumetric packing fraction \cite{bleaney_effective_1941}.  $D_g$  was assumed to be the same as the demagnetization factor for a single particle of the same shape.
% The demagnetizing field can typically only be evaluated analytically in a few cases, such as sample with ellipsoidal shape [ref], infinite cylinders, and infinite sheets [ref], where the internal magnetic field in the material is uniform when subjected to a uniform external field.
 A numerical calculation confirmed this model is a good approximation within 3\% for a sample of randomly packed spherical particles for  sample aspect ratios $\gamma_g = 0.5$ to $1$, and packing fractions $\phi$ from 0.4 to 0.6 \cite{bjork_demagnetization_2013}.  It remains to be seen how well the prediction holds over a wider parameter range,  in particular at larger $\gamma_g$  where $\chi_{eff}$ is  expected to be larger.

A more general model takes advantage of the fact that exact expressions can be found for homogeneously magnetized ellipsoids of revolution to obtain an expression for ellipsoidal particles homogeneously dispersed in any non-magnetic medium (including suspensions) in which particles are aligned with each other and the external magnetic field \cite{skomski_effective_2007}.  This demagnetization factor is
\begin{equation}    
D = D_p(1-\phi)+D_g\phi 
\label{Eq_demagFactor}
\end{equation}    
\noindent where $D_p$ is the demagnetization factor of each particle (they are assumed to be identical). $D_p$ may be different from the global demagnetization factor $D_g$, and so is an unknown function of particle geometry.
%Eq.~\ref{Eq_demagFactor} may be interpreted as a weighted average of $D_p$ and $D_g$ in $\phi$, where $D_p$ dominates at small $\phi$ and $D_g$ at large $\phi$.  
An expression for $\chi_{eff}$ can be obtained by combining Eq.~\ref{Eq_sus_eff_mat_demag} and \ref{Eq_demagFactor}, which simplifies  if  demagnetization effects are as significant as  they are for typical for single-piece ferromagnetic materials  in the limit where $\chi \gg \chi_{eff}$ to 
\begin{equation}
  \chi_{eff} \approx \frac{\phi}{D_p(1-\phi)+D_g\phi}.
  \label{Eq_sus_model}
\end{equation}     
\noindent  To our knowledge, it has not yet been tested whether this model captures the effects of different particle shapes on $\chi_{eff}$ -- specifically there is no model or data on how $D_p$ depends on particle aspect ratio or other parameters.

%divergence at phi_c
An alternate model designed for the limit of high packing fraction $\phi$ assumes that  magnetic field lines tend to go from one ferromagnetic particle to another along regions of high susceptibility, and thus concentrate their density in paths along the shortest distances between particles \cite{martin_magnetic_2000}.  It predicts that $\chi_{eff}$ diverges as the gaps between magnetic particles go to zero, approaching the material susceptibility $\chi$, according to
\begin{equation}
  \chi_{eff} = \frac{1}{1- \left(\phi/\phi_c\right)^{1/3}} - 1,
  \label{Eq_Martin}
\end{equation} 

\noindent while $\chi_{eff} \ll \chi$.  The critical packing fraction $\phi_c$  physically corresponds to the liquid-solid transition  where  particles with no long-range repulsions just barely touch.  While this model has been tested at low $\phi$, it has not been tested at $\phi$ within 0.08 of the liquid-solid transition where the divergence would be expected to produce  $\chi_{eff} \gg 1$ \cite{martin_magnetic_2000, frick_effective_2002}, so it is not yet known if this divergence can be realized in suspensions.
%{\bf ruled out by bjork and bahl? dry granular materials up to solid transition}

%specific goal
In this manuscript, we test the above predictions for $\chi_{eff}$ for suspensions of cylindrical and spherical particles in cylindrical samples, by measuring $\chi_{eff}$ over a wide range of  packing fraction $\phi$ up to $\phi_c$,  sample aspect ratio $\gamma_g$, and particle aspect ratio $\gamma_p$.   The remainder of the manuscript is organized as follows. We first describe the suspensions used in Sec.~\ref{sec:materials}.  We describe the gradiometer we built to measure $\chi_{eff}$ in Sec.~\ref{sec:gradiometer}, and its calibrations in Secs.~\ref{sec:inductancecal} and \ref{sec:susceptibilitycal}.  We test the linearity of the magnetic response of the suspensions in current and frequency in Sec.~\ref{sec:linearity}. Measurements of $\chi_{eff}$ as a function of $\phi$, $\gamma_g$ and $\gamma_p$ are reported in  Sec.~\ref{sec:parameter_variation}.  We use this to  fit the  demagnetization functions $D_p$ and $D_g$ from Eq. \ref{Eq_demagFactor}  as a function of  aspect ratios $\gamma_p$ and $\gamma_g$,  respectively, in Sec.~\ref{sec:model_test}.  Finally in Sec.~\ref{sec:alignment}, we vary particle aspect ratio $\gamma_p$ in  suspensions of randomly arranged particles to test  whether particle alignment with the magnetic field plays an important  role in in $\chi_{eff}$, an  effect which was not accounted for in Eq.~\ref{Eq_demagFactor}.
%We cover a wider range of $\gamma_g$ than tested before ($1 \le \gamma_g \le 2$ \cite{bjork_demagnetization_2013}), 
%      The model in Eq. \ref{Eq_sus_model} has been tested out by Bjork and Bahl \cite{bjork_demagnetization_2013} for dry spherical particles using numerical simulations for packing fraction range $\phi$ from 0.4 to 0.6, and for 3 sample aspect ratio $\gamma_g$ from 1 to 2. Here in this paper, we test Eq. \ref{Eq_sus_model} for suspensions with different packing fraction $\phi$, sample aspect ratio $\gamma_g$,  and sample aspect ratio $\gamma_g$ than has been covered previously . 

\section{Materials and methods}

\subsection{Materials}
\label{sec:materials}

  We suspended iron particles (density 7.834 $kg/m^3$ and purity 99.5\%) of mean diameter 29 $\mu m$, where 90\% of particle diameters are within the range 18-40 $\mu m$, purchased from Chemicalstore.com. The particles are  nearly spherical with a standard deviation of 4\% in the diameter. We use these nearly spherical particles in experiments unless otherwise specified. The suspending liquid was a eutectic alloy of gallium and indium known as eGaIn,  which was produced as described in \cite{CBCVB17}.  We used a liquid metal for its potential in magnetohydrodynamic applications.  The properties of the liquid metal are not expected to be important here,  other than the effect of its conductivity $3.40\times10^6$ S/m \cite{DCLWWW08} contributing to stronger eddy currents that could reduce $\chi_{eff}$ at high frequencies of  applied alternating magnetic field (see Sec.~\ref{sec:linearity}).  Samples were kept in an acid bath to prevent oxidation of the metals \cite{CBCVB17}.  Dry granular samples were obtained by mixing the iron particles with non-magnetic sand.  
  
In either case, the packing fraction $\phi$ was obtained by measuring masses of the constituent materials, and using density to convert to a packing fraction by volume of the magnetic material (iron) divided by the total volume taken up by the sample.  In the case of dry granular samples, the total volume of the sample was measured directly as the volume taken up in the sample container, which includes some air.

\subsection{Experimental setup}
\label{sec:gradiometer}

\begin{figure}
\centering
\includegraphics[width=3.4in]{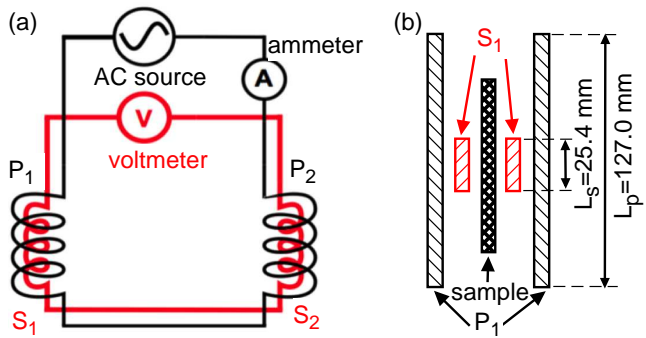} 
\caption{(a) The circuit diagram of the gradiometer used to measure the effective susceptibility $\chi_{eff}$ of a sample based on the change in mutual inductance of a solenoid pair. (b) A diagram (not to scale) that shows a cross-section of the coils and sample. }
\label{Fig_gradiometer}
\end{figure} 

%theory
Measurements were taken using a gradiometer setup which consists of two pairs of inductor coils shown in Fig.~\ref{Fig_gradiometer}.  The gradiometer measures $\chi_{eff}$ of a material sample based on how it changes the mutual inductance between two surrounding coils $P_1$ and $S_1$. The sample sits in a cylindrical tube which is placed inside the secondary coil $S_1$, while coil $S_1$ is inside the primary coil $P_1$.   There is another nominally identical set of coils $P_2$ and $S_2$.  An alternating current $I_p$ is applied at frequency $f$ (angular frequency $\omega=2\pi f$) through the primary coils $P_1$ and $P_2$, while the induced voltage $\epsilon_{ind}$ is measured across both the secondary coils $S_1$ and $S_2$.   The secondary coils are linked in the opposite direction in the circuit such that the mutual inductances of each pair of primary and secondary coils -- $M_1$ and $M_2$, respectively -- cancel in their contribution to the measured $\epsilon_{ind}$ when there is no sample inside coil $S_1$.  In ideal theory, the induced voltage is then proportional to $\chi_{eff}$.  In practice, the two pairs of coils are not identical which we account for with a small correction factor $\Delta M = M_1 - M_2$.    Furthermore, there is a background voltage  noise $\epsilon_{noise}$ measured when there is no sample and no applied current.  The theory of induction allows derivation of an expression for the induced voltage $\epsilon_{ind}$   
\begin{equation}
  \epsilon_{ind}^2 = \epsilon_{noise}^2 + \omega^2 I_p^2 [\alpha M_1 \chi_{eff} + \Delta M]^2 \ .
  \label{Eq_sus_meas}
\end{equation} 
\noindent  where $\alpha$ is the fraction of volume of coil $S_1$ filled by the sample. This expression assumes that the background noise is distributed among all phases, which differs from  the fixed phase of the induction signal, so that the root-mean-square values of the their respective contributions to the induced voltage are added in quadrature.

%dimensions 
The  geometric parameters of the system are as follows.  The primary coil $P_1$ has length $L_p = 127.0 \pm 0.2$ mm, diameter $d_p = 50.8 \pm 0.2$ mm, and $N_p = 332 \pm 22$ turns of wire.  The secondary coil $S_1$ has length $L_s = 25.4 \pm 0.2 $ mm, diameter $d_s = 14.8 \pm 0.2$ mm, and $N_s = 190 \pm 14$ turns of wire. The coils $S_2$ and $P_2$ are nominally identical to $S_1$ and $P_1$, respectively.   All samples were prepared in cylindrical containers of length $L$ that satisfies $L_s < L < L_p$ so that  they were fully contained in the  uniform region of the applied field and their edge effects have a  minimal effect on the flux seen by coil $S_1$.  The filling fraction of the coil $S_1$ is then given by $\alpha=d^2/d_s^2$, where $d$ is the diameter of the sample.   We aligned the sample vertically within coil $S_1$ by finding the position of maximum measured induced voltage, as misalignment along the axis of the cylinder results in a reduced signal. The samples had inner diameter $d=10.2 \pm 0.1$ mm unless otherwise specified, for aspect ratio corresponding to a typical filling factor $\alpha=d^2/d_s^2=0.471$. The sample aspect ratio is given by $\gamma_g=L/d$.

%errors and typical values 
Here we summarize some typical electrical measurement parameter values and errors.   We report  root-mean-square values for all of our measurements of both alternating current and voltage throughout the paper.   For our measurements, the applied alternating current is typically $I_p=65 \pm 0.5$ mA  (corresponding to an 0.8\% error) unless otherwise noted, where the error is given by the manufacturer (Agilent model 34401A multimeter).  We typically report measurements at frequencies $f$ ranging from 200 to 2000 Hz, and $\chi_{eff}$ is calculated from Eq.~\ref{Eq_sus_meas}, using an unweighted average over this frequency range unless otherwise specified.  At these typical measurement values and when $\chi_{eff} \ge 1.2$, for example, we measure $\epsilon_{ind} \ge 23$ mV for different samples, with an uncertainty $\le 0.2$\% ($\ge0.05$ mV) based on the $0.06\%\epsilon_{ind} + 0.04$ mV  reported by the manufacturer, which  is generally less than the uncertainty on the current measurement.   The noise term $\epsilon_{noise}$ is due to electronic noise, and as such, varies when the measurement equipment is on.  It is thus measured as $\epsilon_{ind}$ at a weak signal with frequency $f=5$ Hz at $I_p=65$ mA.   We measured $\epsilon_{noise} = 3$ mV on average, with a standard deviation of 0.4 mV over the course of a series of experiments shown in one plot, or 1 mV over the longer time scale of different measurement series.     When added in quadrature as in Eq.~\ref{Eq_sus_meas}, this leads to an error on $\epsilon_{ind}$ of less than  0.2\% for $\chi_{eff} \ge 1.2$ for example,  which is small compared to the other errors for these typical measurement parameters.  This error becomes dominant when the signal is smaller, notably where we test the linearity of the signal at small values of $I_p$ or $f$ in Sec.~\ref{sec:linearity}, or small $\phi$ where $\chi_{eff} \ll 1$. 
%For example the error on $\chi_{eff}$ is 6\% at the smallest packing fraction we measured in this paper $\phi=4.3$\% where $\chi_{eff}=0.18$, and is 4\% at next packing fraction point $\phi=5$\% where $\chi_{eff}=0.30$. 
Similarly, the absolute error on $\chi_{eff}$ from the error of $\Delta M$ is 0.01, or equivalently less than 0.8\% of $\alpha M_1\chi_{eff}$ when $\chi_{eff} \ge 1.2$ for the typical measurement parameters (see Sec.~\ref{sec:inductancecal} on how values of $M_1$ and $\Delta M$ are obtained).  Thus, the largest systematic source of error in calculating $\chi_{eff}$ from Eq.~\ref{Eq_sus_meas} unless otherwise noted typically comes from the 0.8\% on the applied current $I_p$ for our typical parameters and $\chi_{eff} \ge 1.2$.

When we repeated measurements by turning off the electronics, taking a sample container out from inside the coils, putting the sample back, and turn on the electronics again, the run-to-run standard deviation was 2.5\% for suspensions and granular samples, and 0.2\% for macroscopic solid pieces.   The larger run-to-run variation of suspensions and powders may  come from the rearrangement of particles as the sample containers are disturbed, but it is smaller than the 6\% standard deviation observed in numerical simulations \cite{bjork_demagnetization_2013}.

\subsection{Inductance calibration}
\label{sec:inductancecal}

\begin{figure}
\centering
\includegraphics[width=3.4in]{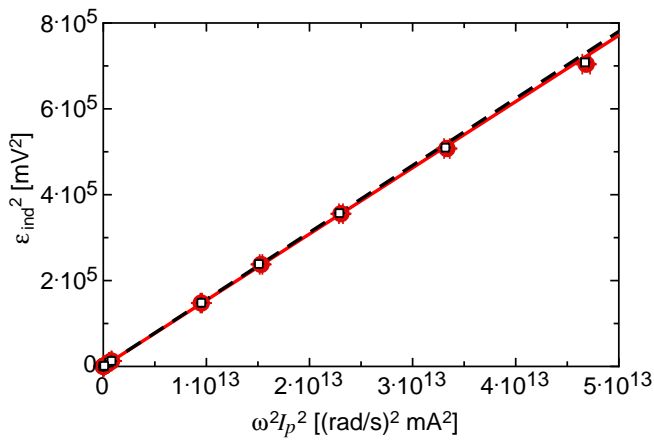}  
\caption{Induced voltage from the mutual inductance of each pair of solenoid coils in isolation.  Solid circles:  coils $P_1$ and $S_1$.  Open squares:  coils $P_2$ and $S_2$. The slopes of the fits yield the squares of the mutual inductances $M_1$ (solid line) and $M_2$ (dashed line), used for calibration of the apparatus.}
\label{Fig_mutual}
\end{figure}  

To provide calibration values of $M_1$ and $\Delta M$ in Eq. \ref{Eq_sus_meas}, we measure the mutual inductances of each coil independently, in each case removing the other coil from  the circuit and measuring without a sample.  In these cases, the measured  voltage is expected to be 
\begin{equation}
\epsilon_{ind}^2 = \epsilon_{noise}^2 + \omega^2 I_p^2 M_i^2,
\label{Eq_mutual}
\end{equation} 
\noindent where $i=1$ or $2$ is the coil pair index number. Measurements of $\epsilon_{ind}^2$ are shown as a function of $\omega^2 I_p^2$ in Fig.~\ref{Fig_mutual} for both coil pairs. We fit a linear function plus a constant to each to obtain the slopes $M_1$ and $M_2$, respectively.  The error bars in the figure represent the sum of a 0.2 \% standard deviation of multiple repetitions and a 0.8\% systematic error. To obtain a fit with a reduced Chi-squared of 1 (where the reduced Chi-squared value of a fit  corresponds to the mean-square difference between the data and fit, normalized by the error), we adjust the percentage input errors to 1.6\% and 1.9\% for coil pairs 1 and 2, respectively. The fit yields $M_1=(1.245 \pm 0.004) \times 10^{-4}$ H and $M_2=(1.253 \pm 0.004) \times 10^{-4}$ H. These measured values are consistent with the expected theoretical value $M = N_pN_sA_s/L_p=(1.2 \pm 0.1) \times 10^{-4}$ H based on the dimensions of the setup, where $A_s=\pi d_s^2/4$ is the cross-sectional area of the secondary coil. The difference between these measured mutual inductances is $\Delta M = M_2-M_1 = (8 \pm 6) \times 10^{-7}$ H.  These values $M_1$ and $\Delta M$ are used as calibrations to  calculate $\chi_{eff}$ from Eq.~\ref{Eq_sus_meas}.

\subsection{Susceptibility calibration}
\label{sec:susceptibilitycal}

\begin{figure}
\centering
\includegraphics[width=3.4in]{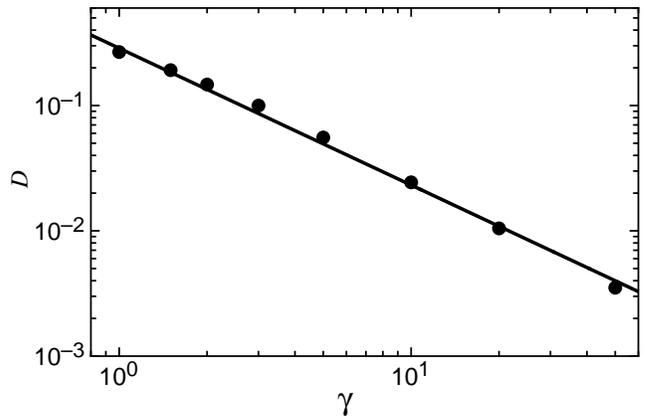}  %  {eg article/figure1.eps}
\caption{Demagnetization factor $D$ for a single-piece cylindrical sample as a function of aspect ratio.  The data is reproduced from Chen et al.~\cite{chen_fluxmetric_2006}. A power law is fit to obtain a reference curve to account for the demagnetization effect in our measurements.  }
\label{Fig_demagFactor}
\end{figure}  

We used single-piece solid cylindrical samples to calibrate $\chi_{eff}$ measurements in our setup. To account for the demagnetization effect, we use for  reference a numerical simulation of the demagnetization factor $D$ for single-piece cylindrical samples of various aspect ratios $\gamma$ from Chen et al.~\cite{chen_fluxmetric_2006}, shown in Fig.~\ref{Fig_demagFactor}. We fit the function
\begin{equation}
  D = A\gamma^{n}.
  \label{Eq_demagFactor_Chen} 
\end{equation}
\noindent to this data, over the range $0.7<\gamma<50$, which covers our measurement range. We adjusted the input error to be 8\% to obtain a reduced Chi-squared of 1, yielding $A=0.31 \pm 0.01$ and $n=-1.12 \pm 0.02$.

\begin{figure}
\centering
\includegraphics[width=3.4in]{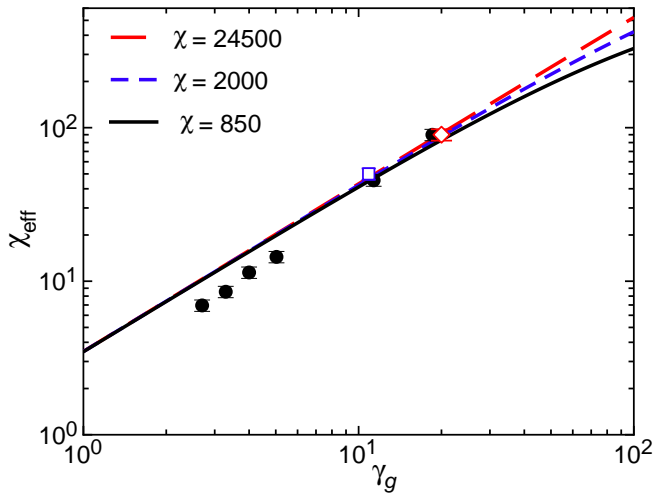} 
\caption{Measured effective susceptibility $\chi_{eff}$ of single-piece solid samples as a function of aspect ratio $\gamma$.  Solid circles: material susceptibility $\chi=850$. Open square: $\chi=2000$.  Open diamond:  $\chi=24500$. Lines:  fits of numerical simulation results from Chen et al.~\cite{chen_fluxmetric_2006}, for the different  material susceptibilities $\chi$ as given in the legend to obtain reference curves $\chi_{ref}$.
}
\label{Fig_solidRod}
\end{figure} 

\begin{figure}
\centering
\includegraphics[width=3.4in]{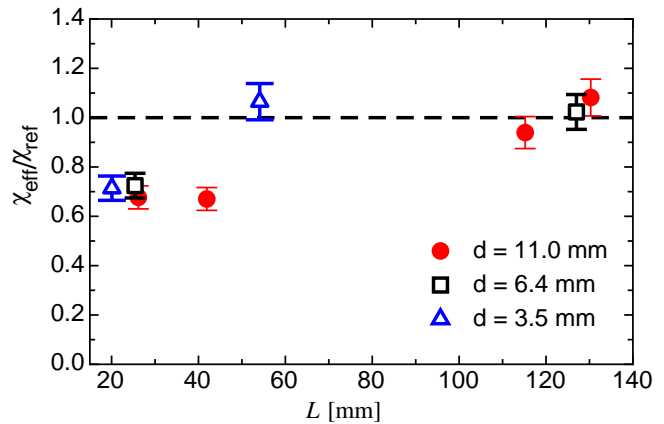}  
\caption{Ratio between the measured effective susceptibility $\chi_{eff}$ for single-piece solid samples and the reference value from Chen et al.~\cite{chen_fluxmetric_2006}, as a function of sample length $L$. Sample diameters $d$ are indicated in the legend.  We adjust measurements of $\chi_{eff}$ for suspensions in later plots for $L < 50$ mm by a calibration factor based on this ratio. 
}
\label{Fig_cali_length}
\end{figure}

 To calibrate our setup, we measured $\chi_{eff}$ of single-piece solid samples with different dimensions (the specific lengths and diameters are indicated in Fig.~\ref{Fig_cali_length}) and materials. Measured $\chi_{eff}$ are shown in Fig.~\ref{Fig_solidRod} at different aspect ratios $\gamma$ for ferrite ($\chi=850$, Fair-Rite Products Corp.),  mu-metal ($\chi=2000$, Aperam), and Permalloy ($\chi=24500$, National Magnetics Group, Inc.).   The plotted errors are the sum of the 0.2\% run-to-run variation and 0.8\% systematic error.   A reference curve $\chi_{ref}$ is shown in Fig.~\ref{Fig_solidRod} for each material, which is calculated by inserting the fit function for $D$ (Eq.~\ref{Eq_demagFactor_Chen}) into Eq.~\ref{Eq_sus_eff_mat_demag} with $\phi=1$.   Since these $\chi$ are all much greater than 1, the  predicted $\chi_{ref}$ curves are all close to each other.  For aspect ratio $\gamma \ge 10$, the measured $\chi_{eff}$ values collapse onto the reference curves $\chi_{ref}$ within a root-mean-square difference of 7\%. However, for  $\gamma\le5$ the measured $\chi_{eff}$ are about 30\% smaller than the reference curve. 

To come up with an appropriate calibration adjustment, we first consider that sample aspect ratio may not be the primary parameter which it could depend on.  In the ideal theory assumed in Eq.~\ref{Eq_sus_meas}, if $L_s \ll L \ll L_p$, the magnetic field inside $S_1$ is expected to be uniform.  In practice,  fringe effects may add a correction. To come up with a calibration adjustment as a function of the sample length $L$, we replot our measurements of $\chi_{eff}$ for single-piece samples from Fig.~\ref{Fig_solidRod} normalized by the reference curve $\chi_{ref}$ as a function of the sample length $L$ in Fig.~\ref{Fig_cali_length}.  Different sample diameters $d$ are indicated in the figure legend with uncertainties of 0.2 mm. A  systematic dependence on $L$ is observed in Fig.~\ref{Fig_cali_length},  similar to the trend in Fig.~\ref{Fig_solidRod}.  In contrast, there is no systematic trend in $d$, as some points for each value of $d$ are in each of the lower and upper ranges of $\chi_{eff}/\chi_{ref}$.   This confirms the calibration should be made as a function of $L$, but not as a function of $d$.  For $L \ge 54$ mm, the reference curve agrees with our  measurement  within a root-mean-square difference of 7\% (a  7\% error bar is plotted in Fig.~\ref{Fig_cali_length} to see this). 
However, for $L \le 42$ mm ($=1.7L_s$), $\chi_{ref}$ is an average of $40\% \pm 4\%$ larger than $\chi_{eff}$.  Based on these results, we introduce a calibration factor in which the following measurements for $\chi_{eff}$ are shifted upwards by  a factor of 1.4 for samples with $L \le 42$ mm.  We note that  most of our  samples in later measurements have $L\ge54$ mm, and this calibration factor only needs to be applied to a few of our shortest samples,  specifically for aspect ratio $\gamma_g=2.5$ in Fig.~\ref{Fig_suscept_phi}, samples with $\gamma_g<5$ in Fig.~\ref{Fig_suscept_vs_globalAspect}, and samples with aspect ratio $\gamma_g=4.1$ in Figs.~\ref{Fig_suscept_vs_gammaP_wire} and \ref{Fig_suscept_vs_phi_wire}. Based on the variation of 7\% around the reference curve observed here, we also introduce an error of 7\% from unknown sources when comparing samples in all following measurements of $\chi_{eff}$.

\section{Results}

\subsection{Linearity  of  magnetic response}
\label{sec:linearity}

\begin{figure}
\centering
\includegraphics[width=3.4in]{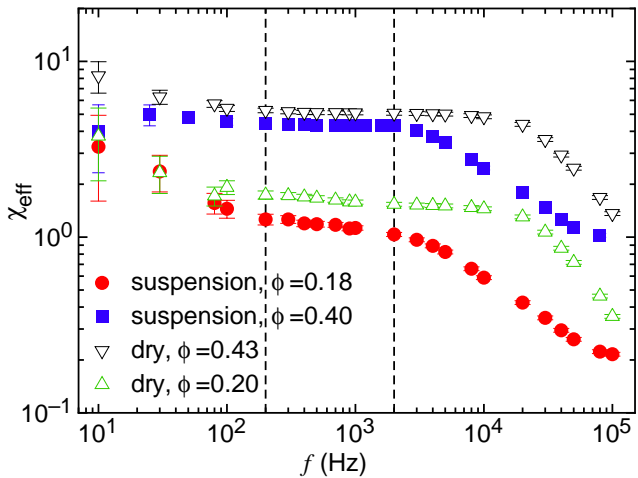}  %  {eg article/figure1.eps}
\caption{Examples of $\chi_{eff}$ as a function of frequency $f$.  Solid symbols: suspensions of iron particles in eGaIn at $\phi=18\%$ (circles) and $\phi=40\%$ (squares). Open symbols:  dry granular materials at $\phi=20\%$ (up-pointing triangles) and $\phi=43\%$ (down-pointing triangles).   $\chi_{eff}$ reaches a plateau for  $f<2000$ Hz.  The vertical lines indicate the bounds of the frequency range where $\chi_{eff}$ is averaged over for measurements  reported in other plots. 
}
\label{Fig_suscept_vs_freq}
\end{figure} 

To  test the linearity of the magnetic properties of the materials with frequency, some examples of the measured susceptibility $\chi_{eff}$ are shown  for dry granular materials and suspensions as a function of frequency $f$ in Fig.~\ref{Fig_suscept_vs_freq}, at sample aspect ratio $\gamma_g = 11$, length $L=112.2$ mm, and packing fractions $\phi$  shown in the legend. The error bars plotted the quadratic sum of the 2.5\% run-to-run standard deviation and the 0.4 mV error on the noise voltage measurements, the latter of which tends to lead to a large error at low frequencies where the signal is weak.  A plateau in $\chi_{eff}$ is found at frequencies $f<2000$ Hz for all suspensions of nearly spherical particles  reported in this paper.  At higher frequencies, $\chi_{eff}$ decreases,  qualitatively similar to the frequency response of other magnetic materials.  The  decrease starts at lower frequencies for suspensions than dry granular materials, which may be expected due to stronger eddy currents in the higher conductivity suspensions.   
%{\bf mention  in discussion? due to the difficulty of reorienting  magnetic dipoles at high frequency. Although in suspensions, an additional degree of freedom exists as the particles can rotate in the liquid --  this would predict opposite trend seen in fig.}.  
At frequencies $f<200$ Hz,  the data remain consistent with the plateau, however there are large relative uncertainties in this range due to the low voltage signal.  Thus, in other plots in this paper, we report the averaged $\chi_{eff}$ over the range of 200 Hz to 2000 Hz as the representative value for the low-frequency plateau, unless we specify otherwise that we found the low-frequency plateau in a different range.   This could introduce an error if there is a trend in $\chi_{eff}$ with frequency, as seen for $\phi=18\%$ suspension in Fig.~\ref{Fig_suscept_vs_freq}.  In this case, which is comparable to the worst case, using the mean of $\chi_{eff}$ for frequencies in the range of 200 Hz to 2000 Hz can underestimate a fit in the zero-frequency limit by up to 3\%, which is negligible compared to the 7\% error we use when comparing samples.

\begin{figure}
\centering
\includegraphics[width=3.4in]{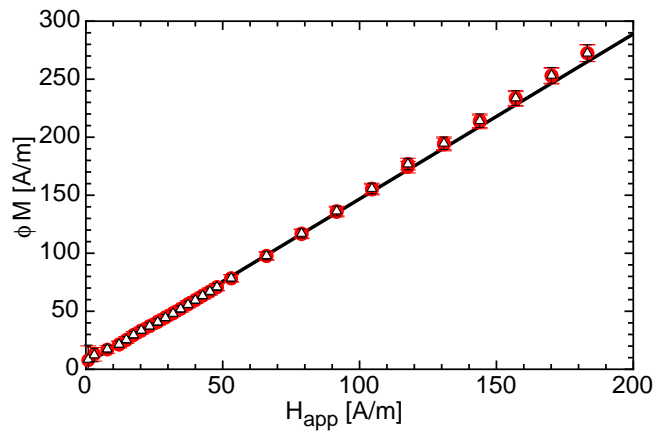}  %  {eg article/figure1.eps}
\caption{Magnetization  per unit volume of sample $\phi M = \chi_{eff}H_{app}$ of suspensions as a function of applied magnetic field $H_{app}$. Closed circles:  increasing $H_{app}$ (or $I_p$).  Open triangles: decreasing $H_{app}$ (or $I_p$). Line: linear fit.  The suspension behaves as a linear paramagnetic material, with no hysteresis or significant remnant magnetization.}
\label{Fig_paramag}
\end{figure} 
 
We next  test whether the magnetic response is linear in the applied magnetic field $H_{app}$ (equivalently, whether $\chi_{eff}$ is  independent of $H_{app}$), and whether the suspensions behave more like paramagnetic or ferromagnetic materials.   We plot the magnetization per unit volume of sample $\phi M = \chi_{eff}H_{app}$ vs.~the applied magnetic field $H_{app}=I_p N_p/L_p$ in Fig.~\ref{Fig_paramag} for a suspension with $\phi=0.34$, $\gamma_g=2.5$, and $L= 25.40$ mm.   We performed these measurements with histories of both increasing and decreasing applied current $I_p$ ($\propto H_{app}$).  It is seen in Fig.~\ref{Fig_paramag} that these ramps give  equivalent results,  indicating a lack of hysteresis  in the measured range.  To test the linear response, we fit a linear function with a constant offset to these data where the random error is the quadratic sum of the 2.5\% run-to-run standard deviation and the 0.4 mV random error on voltage measurements, which yields a reduced Chi-squared of  0.8.    The consistency of the linear fit with the data confirms the data are consistent with a $\chi_{eff}$  independent of $H_{app}$  over this range,  verifying the linearity assumed in deriving Eq.~\ref{Eq_mutual}. The error bars plotted in Fig.~\ref{Fig_paramag} include both these systematic and random errors.  The constant offset in the linear fit was 5 A/m,  which is consistent with $M=0$ at $H=0$ within the error of 12 A/m on $M$ at that point due mainly to the 0.5 mA systematic error on the current,  so there is no  resolvable remnant magnetization.  These properties suggest the suspensions  behave as linear paramagnetic materials in this range, which is simpler for both modeling and control, despite the fact that the particles themselves are ferromagnetic.   The suspensions are known behave like a ferrofluid such that after an applied magnetic field is removed, the particles separate and flow like a liquid \cite{CBCVB17}.  This allows the ferromagnetic particles to move around in the liquid and reorient more freely than magnetic domains in a solid  to avoid hysteresis and remnant magnetization.

\subsection{Variation of $\chi_{eff}$ with aspect ratios and packing fraction}
\label{sec:parameter_variation}

%motivation
Now that we have calibrated the apparatus and established linearity of the response over our measurement range,  we now measure the  dependence of the effective susceptibility $\chi_{eff}$ on the sample packing fraction $\phi$, sample aspect ratio $\gamma_g$, and particle aspect ratio $\gamma_p$, to test and fit the model predictions of Eq.~\ref{Eq_sus_model}. 
       
\begin{figure}
\centering
\includegraphics[width=3.4in]{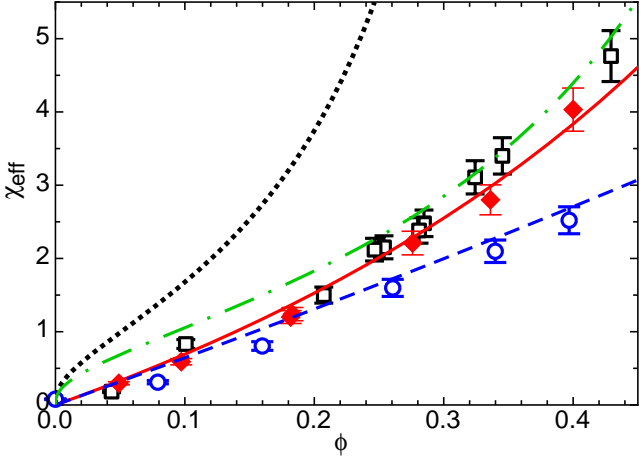}  %  {eg article/figure1.eps}
\caption{Effective susceptibility $\chi_{eff}$ as a function of packing fraction $\phi$ for spherical particles ($\gamma_p=1$).   Squares:  dry granular material, $\gamma_g=11$.  Diamonds:  suspension, sample aspect ratio $\gamma_g=11$.  Circles: suspension, $\gamma_g=2.5$.   Suspensions exhibit a slightly smaller $\chi_{eff}$ than dry granular materials.  Lines: fit of Eq.~\ref{Eq_Chi_A_n} for $\gamma_g=11$ (solid line) and $\gamma_g=2.5$ (dashed line), where fit parameters  are obtained from simultaneously fitting data of Figs.~\ref{Fig_suscept_phi}, \ref{Fig_suscept_vs_globalAspect}, and \ref{Fig_suscept_vs_particleAspect_brokenrods}.  Eq.~\ref{Eq_Martin} is shown for $\phi_c=74\%$ (dashed-dotted line) and the measured liquid-solid transition $\phi_c=40.7\%$ (dotted line).  The divergence  at the liquid-solid transition predicted by Eq.~\ref{Eq_Martin}  is not  observed in the measurements.
}
\label{Fig_suscept_phi}
\end{figure} 

      %phi-dependence
 Figure \ref{Fig_suscept_phi}  shows how $\chi_{eff}$ varies with packing fraction $\phi$ for two series  of suspensions and one of  dry granular materials  with spherical particles ($\gamma_p=1$), and  sample aspect ratios $\gamma_g=2.5$ and 11.  Data at $\gamma_g=2.5$ have been  shifted upwards by a factor of 1.4  according to the calibration in Fig.~\ref{Fig_cali_length}.  Measurements are made at packing fractions up to the liquid-solid transition  $\phi_c=40.7 \pm 0.3\%$ for the suspension,  defined as the lowest packing fraction where a non-zero yield stress is measured.   Measurements of the yield stress for these samples were reported in a previous paper \cite{CBCVB17}.  For  each series, $\chi_{eff}$ increases with increasing $\phi$.  On average, $\chi_{eff}$ of dry  granular materials is higher than that of suspensions by 11\% at the same $\phi$.

\begin{figure}
\centering
\includegraphics[width=3.4in]{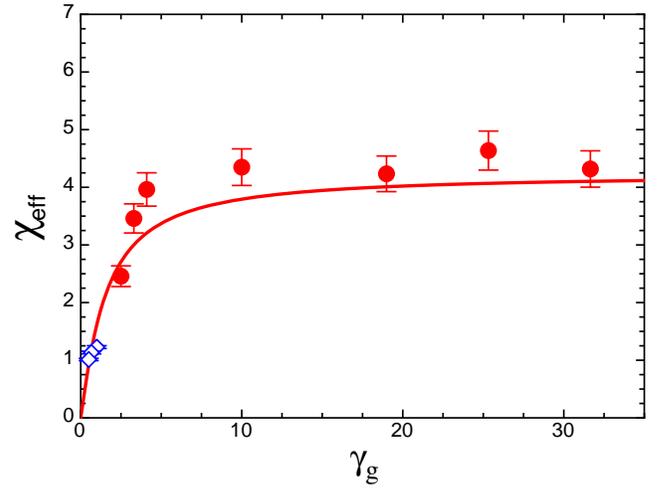}  %  {eg article/figure1.eps}
\caption{Effective susceptibility $\chi_{eff}$  of suspensions as a function of sample sample aspect ratio $\gamma_g$. Solid symbols: spherical particles ($\gamma_p=1$) at $\phi=40\%$. Line: model result of Eq. \ref{Eq_Chi_A_n}, where fit parameters  are obtained from simultaneously fitting data of Figs.~\ref{Fig_suscept_phi}, \ref{Fig_suscept_vs_globalAspect}, and \ref{Fig_suscept_vs_particleAspect_brokenrods}. Open Symbols: numerical simulation of spherical particles at $\phi=40\%$\citep{bjork_demagnetization_2013}.
}
\label{Fig_suscept_vs_globalAspect}
\end{figure} 

%\gamma_g
      Figure \ref{Fig_suscept_vs_globalAspect} shows the effective susceptibility $\chi_{eff}$ as a function of sample aspect ratio $\gamma_g$, for spherical particles ($\gamma_p = 1$) and $\phi=40\%$. To vary $\gamma_g$  and satisfy the condition $L_s < L < L_p$, the sample diameter $d$  had to be varied along with the length.  We use $d=10.2 \pm 0.1$ mm for $\gamma_g < 15$, $d=7.1 \pm 0.1$ mm for $\gamma_g = 19$, and $d=3.7 \pm 0.1$ mm for $\gamma_g > 22$. Measurements for  $\gamma_g < 5$ have been  adjusted upward by 40\%  according to the calibration in Fig.~\ref{Fig_cali_length}. At small $\gamma_g$,  $\chi_{eff}$ increases with $\gamma_g$, and reaches a plateau for $\gamma_g \stackrel{>}{_\sim} 10$.          For  comparison to previous work in Fig.~\ref{Fig_suscept_vs_globalAspect}, we show numerical simulation results of randomly packed spherical particles at $\phi=40\%$ \citep{bjork_demagnetization_2013} (Bjork et al.~reported demagnetization factors $D$ \citep{bjork_demagnetization_2013}, which we converted to $\chi_{eff}=\phi / D$.  Their aspect ratio  was defined as the inverse of our aspect ratio definition.). The simulation data follow the same trend as ours.

\begin{figure}
\centering
\includegraphics[width=3.4in]{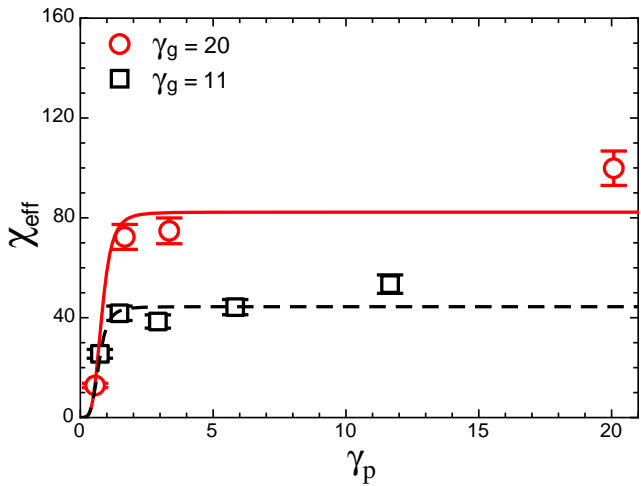}  %  {eg article/figure1.eps}
\caption{Effective susceptibility $\chi_{eff}$ of suspensions of cylindrical rods forced to be aligned with the applied magnetic field as a function of particle aspect ratio $\gamma_p$, with $\phi\approx100\%$. Values of $\gamma_g$ are given in the legend.  Lines: model result of Eq. \ref{Eq_Chi_A_n} for $\gamma_g=11$ (dashed line) and $\gamma_g=20$ (solid line), where fit parameters  are obtained from simultaneously fitting data of Figs.~\ref{Fig_suscept_phi}, \ref{Fig_suscept_vs_globalAspect}, and \ref{Fig_suscept_vs_particleAspect_brokenrods}.   The simultaneous fits here and in Figs.~\ref{Fig_suscept_phi} and \ref{Fig_suscept_vs_globalAspect} confirm the validity of Eq.~\ref{Eq_Chi_A_n} within a root-mean-square difference of 17\%.
}
\label{Fig_suscept_vs_particleAspect_brokenrods}
\end{figure}

    %\gamma_p
      To test the  dependence of $\chi_{eff}$ on particle aspect ratio $\gamma_p$ in  Eq.~\ref{Eq_sus_model} which assumes that particles are aligned with the applied magnetic field, we made dry samples of stacked cylindrical particles where the particles were forced to be aligned with the applied magnetic field.  To make such aligned samples while holding $\gamma_g$ and $\phi$ constant, we cut a 130 mm long cylindrical ferrite rod into collections of gradually smaller pieces to obtain a series of decreasing $\gamma_p$. Each piece was nearly cylindrical, with roughness on a scale of 1 mm at the two ends of the cylinder due to the cutting process. The packing fraction $\phi$ ranged from 100\% to 97\% due to some  loss of material.  This resulted in a number of pieces ranging from 1 for the largest $\gamma_p$ to 32 for the smallest $\gamma_p$ of the series.   The pieces were  arranged in a  stack in the sample container with a common cylindrical axis aligned with the applied magnetic field.  The measured $\chi_{eff}$ as a function of $\gamma_p$ for these aligned particles is shown in Fig.~\ref{Fig_suscept_vs_particleAspect_brokenrods} for two series, one with $\gamma_g= 20$ and one with $\gamma_g=11$. $\chi_{eff}$ initially increass with increasing $\gamma_p$ and levels off for larger $\gamma_p$.

\subsection{Testing the models for $\chi_{eff}$}
\label{sec:model_test}

      The measurements of $\chi_{eff}$ presented in Sec.~\ref{sec:parameter_variation} over a wide range of packing fraction $\phi$, sample aspect ratio $\gamma_g$, and particle aspect ratio $\gamma_p$  now allow us to test the model of Eq.~\ref{Eq_sus_model}.
% Figs. \ref{Fig_suscept_phi}-\ref{Fig_suscept_vs_particleAspect_brokenrods}. 
%Eq. \ref{Eq_demagFactor_Chen} was established from single cylinderical samples, where aspect ratio $\gamma$ may be considered as either sample asepct ratio $\gamma_g$ or particle asepct ratio $\gamma_p$. Therefore, there is no clear distinction whether the functional form in Eq. \ref{Eq_demagFactor_Chen} should be treated as global effect or local effect for each particles in the suspension. In addition, Eq. \ref{Eq_sus_model} has not specified the form for $D_g$ and $D_p$. 
In Eq.~\ref{Eq_sus_model},  the demagnetization factors $D_g$ and $D_p$ are  unspecified functions of sample and particle geometry, respectively.   While calculations have been made of $D_g$ for some shapes \cite{bjork_demagnetization_2013}, we are not aware of any model for $D_p$.  To fit parameters, we assume that both $D_g$ and $D_p$ follow power laws of the form $D=A\gamma^{n}$ as shown in Fig.~\ref{Fig_demagFactor} for single-piece solid magnets \cite{chen_fluxmetric_2006}.  Inserting this forms into Eq.~\ref{Eq_sus_model} with different fit parameters for $D_g$ and $D_p$ yields our fit function 
\begin{equation}
  \chi_{eff} = \frac{\phi}{A_p\gamma_p^{n_p}(1-\phi)+A_g\gamma_g^{n_g}\phi} \ .
  \label{Eq_Chi_A_n}
\end{equation} 
\noindent We simultaneously least-squares fit all our suspension data in Figs.~\ref{Fig_suscept_phi}, \ref{Fig_suscept_vs_globalAspect}, and \ref{Fig_suscept_vs_particleAspect_brokenrods}   to Eq.~\ref{Eq_Chi_A_n} to obtain the fit parameters $A_g$, $n_g$, $A_p$, and $n_p$.  Input error bars were adjusted to 17\%  to obtain a reduced Chi-squared of 1 from the fit. This indicates the model matches the data within a root-mean-square difference of 17\%.  Plots of Eq.~\ref{Eq_sus_model}  with these fit parameters are shown in Figs.~\ref{Fig_suscept_phi}, \ref{Fig_suscept_vs_globalAspect}, and \ref{Fig_suscept_vs_particleAspect_brokenrods}, where it is seen that the model captures the trends of $\chi_{eff}$ in $\phi$, $\gamma_g$, and $\gamma_p$, respectively.  The corresponding best fit parameters are $A_g=0.4 \pm 0.1, n_g=-1.2 \pm 0.1, A_p=0.16 \pm 0.01$, and $n_p=-4.4 \pm 0.3$. The best fit values of $A_g$ and $n_g$ are consistent with the fit values $A=0.31 \pm 0.01$ and $n=-1.12 \pm 0.02$ from the data of Chen et al.\cite{chen_fluxmetric_2006} in Fig.~\ref{Fig_demagFactor}, confirming that $D_g$ in Eq.~\ref{Eq_sus_model} is consistent with  the demagnetization factor $D$ of individual particles \cite{bjork_demagnetization_2013}.

   %  As can be seen from Fig.~\ref{Fig_suscept_phi} to \ref{Fig_suscept_vs_particleAspect_brokenrods}, the model captures the trend of susceptibility quite well, for different $\phi$, $\gamma_g$, and $\gamma_p$. For example in Fig.~\ref{Fig_suscept_phi}, the model captures the behaviors for sample with different $\gamma_g$ while packing fraction $\phi$ varying from zero to maximum packing fraction $\phi_c$. Although the trend of measured data is captured well by the model, the reduced Chi-square value between model and measurements is 5.28, indicating consistent difference between model prediction and measurement. The r.m.s value between the model and measured data is 4.3. The averaged relative difference between model and measurement is about 12\%.

%Comparison with the model of Martin et al 2000
 %measured the magnetic permeability of a suspension, made of liquid gallium and iron beads. By using beads with diameter 6.35 $mm$, the maximum susceptibility achieved by Martin et al. \cite{martin_magnetic_2000} is about 3.8.

To  compare with the prediction of  Martin et al.~\cite{martin_magnetic_2000}, we plot Eq.~\ref{Eq_Martin} as the dashed-dotted line in Fig.~\ref{Fig_suscept_phi} with $\phi_c=74\%$, the value used by Martin et al.~\cite{martin_magnetic_2000}. The result using this  value of $\phi_c$ happens to match well with our data with sample aspect ratio $\gamma_g=11$. However, no sample aspect ratio dependence was prescribed in Eq.~\ref{Eq_Martin}, and the model does not fit well to data at $\gamma_g=2.5$.  Furthermore, the value of $\phi_c$ suggested by Martin et al.~\cite{martin_magnetic_2000} is unphysically large for a liquid-solid transition of a random arrangement of particles, where the particles just barely touch each other, which was the physical meaning of $\phi_c$ in Martin et al.~\cite{martin_magnetic_2000}.   Our  suspensions have a liquid solid transition at  $\phi_c=40.7\%$,  measured as  the lowest packing fraction where the samples exhibit a non-zero yield stress like a solid \cite{CBCVB17}.    To test the physical intent of that model, we plot Eq.~\ref{Eq_Martin} with $\phi_c=40.7\%$ as the dotted line in Fig.~\ref{Fig_suscept_phi}.   This  prediction greatly  overestimates our measurements, which do not  exhibit the divergence at $\phi_c$ of the prediction.  The lack of an observed divergence in $\chi_{eff}$ in the approach to $\phi_c$ is  similar to simulations of  dry granular materials \cite{bjork_demagnetization_2013}.

\subsection{Effect of particle misalignment}
\label{sec:alignment}

\begin{figure}
\centering
\includegraphics[width=3.4in]{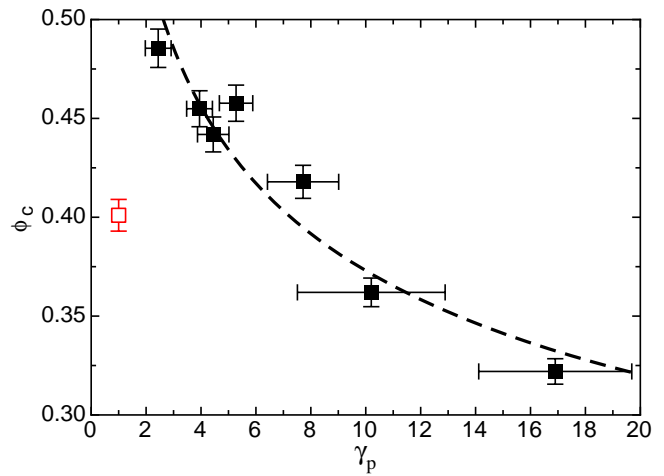}  %  {eg article/figure1.eps}
\caption{Packing fraction of the liquid-solid transition $\phi_c$ as a function of particle aspect ratio $\gamma_p$. Solid symbols: suspensions of cylinders.  Open symbol: suspension of spheres. Dashed line: power law fit to data for cylinders.
}
\label{Fig_wires_packFrac_vs_gammaP}
\end{figure}    

 %particle dimensions
In the previous section,  we tested the model of Eq.~\ref{Eq_sus_model}  for particles aligned with the applied magnetic field,  which was an assumption of the model of Skomski et al.~\cite{skomski_effective_2007}.  However, this is not a very practical case, as real suspensions  of aspherical particles tend to have randomly arranged and oriented particles.  To  characterize how particle misalignment affects $\chi_{eff}$,  we made suspensions of cylindrical particles of various  particle aspect ratios.  We purchased iron wire (Goodfellow) and cut it to make cylindrical particles with different particle aspect ratios $\gamma_p$.  To  obtain samples  with enough particles to avoid significant finite size effects, while minimizing the number of cuts we needed to make,  we used different wire diameters of $0.25$, 0.5, and 1 mm,  for samples with mean particle aspect ratio $\gamma_p > 10$, $5 \le \gamma_p < 10$, and $\gamma_p < 5$, respectively.  For samples of aspect ratio $\gamma_g=4.1$ and length $L=42$ mm, this results in  the ratio of sample diameter to cylinder diameter between 17 and 9, and the ratio of sample diameter to mean particle length between 4.1 and 2.5,  which  is a range where the value of the packing fraction $\phi_c$ of the liquid-solid transition is within 4\% of the infinite-size system limit \cite{DW09, BZFMBDJ10}.    Effects of confinement on alignment are also presumed to be small in this system-size range.  For example, in this range the partial particle alignment from this confinement changes the bulk rheology by less than 3\% \cite{BZFMBDJ10},  but to our knowledge the effect on $\chi_{eff}$ from this confinement has not been characterized.
% The diameter ratios between tube and particles are about 40, 20, and 10 respectively. 

 %packing fraction
 Before we compare $\chi_{eff}$ for particles of different aspect ratio $\gamma_p$, we first identify a meaningful packing fraction criteria for comparison.  Since the packing fraction $\phi_c$ of the liquid-solid transition varies with $\gamma_p$ \cite{DW09, BZFMBDJ10}, it would not be meaningful to compare at the same absolute packing fraction. Rather, we chose to measure $\chi_{eff}$ at a fixed relative packing fraction $\phi/\phi_c$ near the liquid-solid transition to determine the maximum $\chi_{eff}$ we would expect to obtain in the liquid phase for each $\gamma_p$.  Assuming $\chi_{eff}$ increases monotonically with $\phi$ up to $\phi_c$ for any particle shape as seen for spheres in Fig.~\ref{Fig_suscept_phi}, this would be the packing fraction where $\chi_{eff}$ is maximized for each particle shape, and any lower value of $\chi_{eff}$ could be obtained by tuning the packing fraction down to an appropriate value.   $\phi_c$ was measured for each $\gamma_p$ by observing the change in surface reflectivity as the particles poked through the liquid-air interface of the suspension when $\phi>\phi_c$ \cite{brown_shear_2011}. This transition is  sharp and easily observed,  allowing us to measure it with an uncertainty on $\phi_c$ of  $\pm 1\%$.  $\phi_c$ is plotted as a function of particle aspect ratio $\gamma_p$ in Fig.~\ref{Fig_wires_packFrac_vs_gammaP}. The horizontal error bars indicate the standard deviation of particle aspect ratios due to the variation in cut particle lengths.  The measured $\phi_c$ decreases with increasing $\gamma_p$ for cylinders, consistent with previous results \cite{Philipse_Random_Contact_Equation_96}.  For later input into models, a power law is fit to $\phi_c$ for cylinders, yielding $\phi_c=0.62\gamma_p^{-0.22}$.  For comparison we also plot $\phi_c$ for the spherical particles used in earlier sections in Fig.~\ref{Fig_wires_packFrac_vs_gammaP}.  The value of $\phi_c$ for the spheres ford not follow the same trend as the cylinders, not only because of the particle shape, but also the different material source may subtly affect interparticle interactions that can have a significant affect on $\phi_c$ \cite{trappe_jamming_2001}.

\begin{figure}
\centering
\includegraphics[width=3.4in]{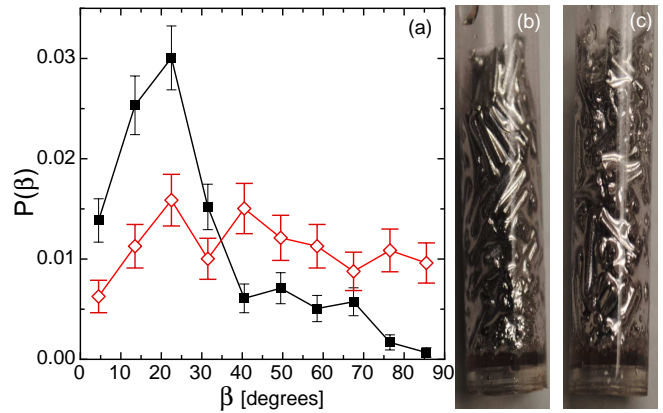} 
\caption{(a) Probability distribution of particle alignment angles $\beta$  relative to the applied magnetic field, for samples with length $L=42$ mm, sample aspect ratio $\gamma_g=4.1$, particle aspect ratio $\gamma_p=5.3$, and packing fraction $\phi/\phi_c=1.02$.  Solid squares: samples were shaken along the axis of the cylindrical  tube to  partly align the particles with the external magnetic field. Open diamonds:  samples were shaken using a vortex mixer,  resulting in a more random alignment.  Pictures  of the samples for the 2 shaking procedures  are shown in panels (b) and (c), respectively, where  the  applied magnetic field is aligned in the vertical direction.}
\label{Fig_PDF_alignment}
\end{figure} 
      
      %shaking
  Since  sample preparation  procedures can affect the alignment of particles,  we also characterize the tendency for the particles to align based on different shaking procedures after the sample was loaded into the cell, but before the magnetic field was applied.   We use a suspension with sample aspect ratio $\gamma_g=4.1$, sample length $L=42$ mm, and particle aspect ratio $\gamma_p=5.3$, at a packing fraction $\phi/\phi_c=1.02$.  This packing fraction is just barely resolvable to be above the liquid-solid transition, so we can observe the alignment of the particles as they poke out the liquid-air interface.  In one case,  samples were shaken  along the axis of the cylindrical tube to  partly align the particles with the external magnetic field, shown in Fig.~\ref{Fig_PDF_alignment}b. In a second case, the same samples were shaken with a combination of linear and rotational shaking (Vortex Genie 2),  resulting in a more random alignment, shown in Fig.~\ref{Fig_PDF_alignment}c  for the same sample parameters.    Note the alignment in the bulk could be quantitatively different from value based on the particles at the boundary, so these pictures at the surface of the sample  should be taken as a coarse characterization of the alignment of particles.  These images were analyzed to measure the angle $\beta$  of each particle relative to the  applied magnetic field, which was aligned with the cylindrical tube axis.   Figure \ref{Fig_PDF_alignment}a shows the  probability distribution  $P(\beta)$ for both of these samples.  For  samples shaken by the vortex mixer, we observe a relatively flat distribution, with a mean $\langle \beta\rangle =44.7^{\circ}$  corresponding to a random alignment. Samples shaken along the axis direction,  while still fairly random,  display a preferred alignment angle $\beta=20^{\circ}$, and a mean $\langle \beta\rangle=27.3^{\circ}$, corresponding to better alignment with the applied magnetic field at $\beta=0^{\circ}$

\begin{figure}
\centering
\includegraphics[width=3.4in]{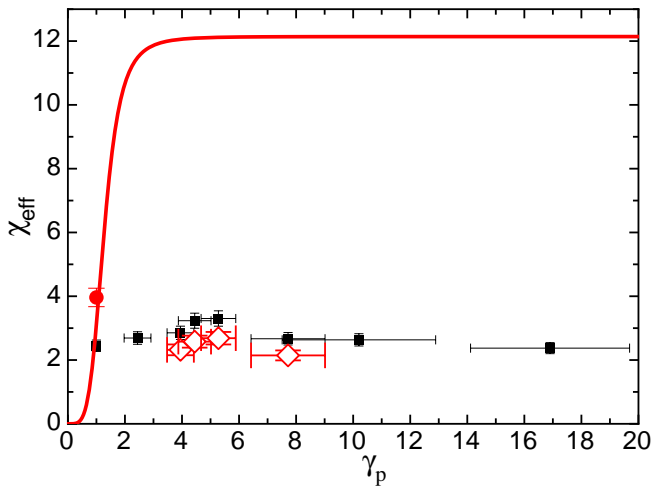} 
\caption{Effective susceptibility $\chi_{eff}$ as a function of particle aspect ratio $\gamma_p$, at $\gamma_g=4.1$ and $\phi/\phi_c=1.02$. Solid squares: samples were shaken beforehand along the axis of the tube to  partly align the particles with the external magnetic field. Open diamonds:  samples were shaken  beforehand using a vortex mixer,  resulting in a more random alignment. Solid circle: random packing of spheres ($\gamma_p=1$), which cannot align.  Solid line: model of Eq.~\ref{Eq_Chi_A_n}. The model overestimates $\chi_{eff}$  by about a factor of 4 for suspensions that are not aligned with the external magnetic field.}
\label{Fig_suscept_vs_gammaP_wire}
\end{figure}     

%chi_eff
Now that we have  identified appropriate packing fractions to compare samples, and characterized the amount of alignment  from different preparation procedures,  we can systematically test trends in effective susceptibility  $\chi_{eff}$  as a function of particle aspect ratio $\gamma_p$ for suspensions of randomly arranged particles.   The measured  $\chi_{eff}$ is shown as a function of  $\gamma_p$ in Fig.~\ref{Fig_suscept_vs_gammaP_wire}, at a fixed relative packing fraction $\phi/\phi_c=1.02$, $\gamma_g=4.1$, $L=42$ mm, and for both shaking procedures.  The error bars on $\gamma_p$ indicate the standard deviation of the aspect ratio due to the distribution of particle lengths in each sample.

 %model
 To test how the model developed for particles aligned with the applied magnetic field applies to randomly arranged particles,  the model prediction of Eq.~\ref{Eq_Chi_A_n} is shown in Fig.~\ref{Fig_suscept_vs_gammaP_wire},  where we use the fit parameter values  obtained from the simultaneous fit of data in Figs.~\ref{Fig_suscept_phi}, \ref{Fig_suscept_vs_globalAspect}, and \ref{Fig_suscept_vs_particleAspect_brokenrods}, and  the power law fit expression for $\phi_c$ from the fit in Fig.~\ref{Fig_wires_packFrac_vs_gammaP} in place of $\phi$ in Eq.~\ref{Eq_Chi_A_n}.  The  model overestimates $\chi_{eff}$  by about a factor of 4  in this case where  the particles are not aligned with the applied magnetic field.  Since that model fit well to data for aligned particles, this indicates the random alignment severely reduces $\chi_{eff}$.  $\chi_{eff}$ is comparable to the value for spheres, which suggests the demagnetization effect may be about as significant for randomly arranged long cylinders as it is for spheres.   The  much larger $\chi_{eff}$ obtained in Fig.~\ref{Fig_suscept_vs_particleAspect_brokenrods}  for large $\gamma_p$ is apparently possible only because the strict alignment of the particles with the  applied magnetic field  reduces  the demagnetization effect.   The trend of higher $\chi_{eff}$ with better aligned particles is  also seen in our  samples with different shaking procedures:  the better aligned particles that were shaken along the cylinder axis had a  consistently 20\% higher $\chi_{eff}$ than  the more randomly arranged  particles that were shaken  by the vortex mixer.   A simple quantitative estimate of the average vector component of  alignment $\cos\langle\beta\rangle$ is also 20\% higher for the  particles shaken along the cylinder axis then those shaken by the vortex mixer.   It suggests,  at least in the ballpark, the decrease of $\chi_{eff}$ in Fig.~\ref{Fig_suscept_vs_gammaP_wire} may be associated with the  change in particle alignment  for these two  samples of randomly arranged particles. However,  extrapolating this simple estimate does not   reach the model of Eq.~\ref{Eq_Chi_A_n},  which suggests that much better alignment would be needed to reach that regime than is likely to be obtained in  suspensions with even partially randomly arranged particles,  regardless of shaking or other procedures used to get a preferential alignment.  
 %With these low values of $\chi_{eff}$ for randomly aligned rods, we find $\chi_{eff}$ to be significantly lower for randomly packed rods than for spheres (we found $\chi_{eff} =4.0$ at the same $\gamma_g$ for spheres in Fig.~\ref{Fig_suscept_vs_globalAspect}), counter to the prediction of  Skomski et al.~\cite{skomski_effective_2007}.

%peak
While  there is little trend in $\chi_{eff}$  over the range of $\gamma_p$  measured  in Fig.~\ref{Fig_suscept_vs_gammaP_wire},  it is notable that $\chi_{eff}$  exhibits a local maximum in $\gamma_p$.   In contrast, Eq.~\ref{Eq_Chi_A_n} predicts $\chi_{eff}$  to be a monotonically increasing function of $\gamma_p$ (as seen in Fig.~\ref{Fig_suscept_vs_particleAspect_brokenrods}).  This local decrease in $\chi_{eff}$ with $\gamma_p$ is not due to the different wire diameters used, as in the range $5 \le \gamma_p < 10$ where $\chi_{eff}$  decreased, the same  diameter wires were used.   Similarly,  finite-size effects cannot explain the peak, as the number of particles   is decreasing over the same range of $\gamma_p$,  which would only be expected to produce  more alignment and a larger $\chi_{eff}$, in contradiction to the trend observed in $\chi_{eff}$.
%On the other hand, in the range $5 \le \gamma_p < 10$ where particle diameter $d=0.5 mm$, the ratio between tube diameter and particle length increases as $\gamma_p$ increases. If the finite-size effect of tube plays a role, one should observe $\chi_{eff}$ increases, since particles should tend to align with tube axis, or magnetic field, better. However, we measured decreasing trend of $\chi_{eff}$ in the range of $5 \le \gamma_p < 10$. It indicates the finite-size effect does not play a role here. 
It could also be proposed that the local peak in $\chi_{eff}(\gamma_p)$  could be due to a competition between the increasing $\chi_{eff}$ in Eq.~\ref{Eq_Chi_A_n} and the decreasing $\phi_c$  with $\gamma_p$.  However, as shown in Fig.~\ref{Fig_suscept_vs_gammaP_wire}, the model of Eq.~\ref{Eq_Chi_A_n} still has no local maximum in this parameter range --  even when accounting for this decreasing $\phi_c$ with $\gamma_p$.  This  insensitivity to $\phi$ in the model is apparent in the limit of large $\gamma_p$ of Eq.~\ref{Eq_sus_model}, which becomes $\chi_{eff} \approx 1/D_g$, independent of $\phi$.  The cause of this  local maximum in $\chi_{eff}(\gamma_p)$ remains unknown.

%packing fractions Dependence of susceptibility  of rods
The data in Fig.~\ref{Fig_suscept_vs_gammaP_wire} were taken at $\phi/\phi_c=1.02$, corresponding  to a jammed state where particles were not free to realign in the applied  magnetic field. If instead particles were at a lower packing fraction in a liquid state, they might be expected to be able to more freely and better align with the applied magnetic field to reach the higher $\chi_{eff}$ predicted by Skomski et al.~\cite{skomski_effective_2007}.  To test this hypothesis, we measured $\chi_{eff}$ as a  function of packing fraction $\phi$,   for 0.5 mm diameter wire cut to length 3.2 mm with a standard deviation of 0.6 mm to obtain a particle aspect ratio $\gamma_p=6.3\pm1.3$, near the peak found in Fig.~\ref{Fig_suscept_vs_gammaP_wire}.  We started with a sample aspect ratio of $\gamma_g=4.1\pm0.3$ at $\phi=0.42$ in a 10.2 mm diameter tube, and diluted the sample with more liquid to increase $\phi$.  The sample aspect ratio decreased to 3.8 as the liquid-solid transition was crossed as the suspension packed more efficiently, without trapped air.  Upon further dilution, the sample aspect ratio increased in inverse proportion to the packing fraction due to the increase in liquid volume. 
 %$\gamma_g=6.7 at $\phi=0.22$.
%$d=10.2\pm01.$ mm
  %For $\phi<\phi_c$, the particles tended to remain clumped together even after after shaking, possibly due to interlocking of the particles.   
 Because the signal was weaker at these lower frequencies, the calibration of $\epsilon_{noise}$ was done with more precision by measuring induced voltage separately before each data point with the current source outputting at the  frequency and applied current of the data point but without a sample.  The  plateau where $\chi_{eff}$ was independent of frequency occurred for $f < 200$ Hz for these cylinders, so the reported $\chi_{eff}$ was obtained from a weighted average of data in that range.  
 
\begin{figure}
\centering
\includegraphics[width=3.4in]{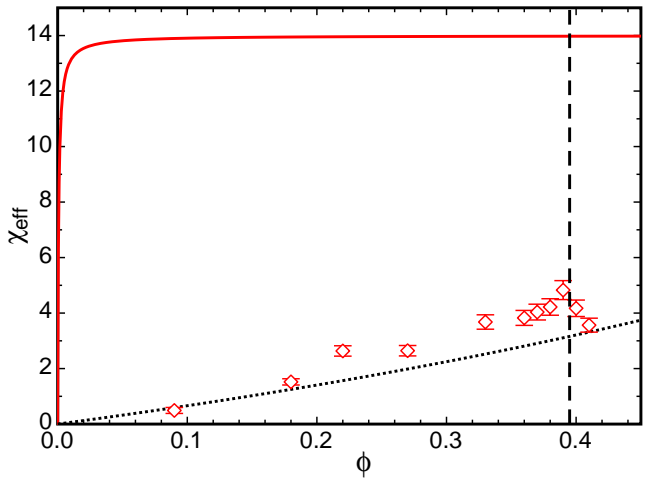} 
\caption{Effective susceptibility $\chi_{eff}$ as a function of packing fraction $\phi$ for particle aspect ratio $\gamma_p=6.3$ cylinders. Solid line:  model of Eq.~\ref{Eq_Chi_A_n} for $\gamma_p=6.3$.  Dotted line: model of Eq.~\ref{Eq_Chi_A_n} for spheres ($\gamma_p=1$). Vertical dashed line: the packing fraction of the liquid-solid transition.  The model still overestimates $\chi_{eff}$ by a factor of 3 or more for suspensions of randomly aligned cylinders, regardless of whether they are in a liquid or solid state.}
\label{Fig_suscept_vs_phi_wire}
\end{figure}     

%results -wires: chi_eff vs phi
Values of $\chi_{eff}$ for these particle aspect ratio $\gamma_p=6.3$ cylinders are shown as a function of packing fraction $\phi$ in Fig.~\ref{Fig_suscept_vs_phi_wire}.  We only report $\chi_{eff}$  for samples shaken in the vortex mixer, as samples shaken vertically to intentionally align particles showed an  increase in $\chi_{eff}$ of typically 20\%, as found in Fig.~\ref{Fig_suscept_vs_gammaP_wire}.  For $\phi \ge 0.33$, the samples had length $L < 42$ mm, so measured $\chi_{eff}$ values were scaled up by a factor of 1.4 according to the calibration in Sec.~\ref{sec:susceptibilitycal}. $\chi_{eff}$ increases with $\phi$ for cylinders as it does for spheres for $\phi<\phi_c$.  We do find a decrease in $\chi_{eff}$ as $\phi$ increases above $\phi_c$, as expected due to the inability of  particles to rearrange for $\phi>\phi_c$.    The prediction of Eq.~\ref{Eq_Chi_A_n} for aspect ratio $\gamma_p=6.3$ is shown as the solid line in Fig.~\ref{Fig_suscept_vs_phi_wire}.  The prediction is again well above the data, by a factor of 3 or more.  A correction for the variation of sample aspect ratio $\gamma_g$ from the dilution according to Eq.~\ref{Eq_Chi_A_n} would not increase $\chi_{eff}$ by more than 10\% for any data point, not nearly enough to match the prediction shown in Fig.~\ref{Fig_suscept_vs_phi_wire}.  For comparison, we also plot the prediction of Eq.~\ref{Eq_Chi_A_n} for aspect ratio 1 as the dotted line in Fig.~\ref{Fig_suscept_vs_phi_wire}.  The aspect ratio $\gamma_p=6.3$ particles do have a slightly higher $\chi_{eff}$ than spheres, and reach up to $\chi_{eff} =4.8$ at the highest packing fraction of the liquid state ($\phi=0.39$).     However, the disagreement with prediction confirms that even in the liquid state, the larger $\chi_{eff}$ predicted by Skomski et al.~\cite{skomski_effective_2007} is not realized, due to to the random arrangement and orientation of particles in suspension which produces a strong demagnetization effect even for large particle aspect ratios.

\section{Conclusions}

In this paper, we  reported measurements of the effective magnetic susceptibility  $\chi_{eff}$ of suspensions as  a function of packing fraction $\phi$, sample aspect ratio $\gamma_g$, and particle aspect ratio $\gamma_p$.  When particles are aligned with the applied magnetic field, the model of Skomski et al.~\cite{skomski_effective_2007}  can be fit with power laws for  the demagnetization factors $D_g$ and $D_p$ describing the aspect ratio dependence of the sample and particles, respectively, in the form of Eq.~\ref{Eq_Chi_A_n} with a root-mean-sqaure difference of 17\%. This was done by simultaneously  varying the three model parameters over the range $0 \le \phi \le 40.7\%$ (up to the liquid-solid transition $\phi_c$), $2.5 \le \gamma_g \le 32$, and $1 \le \gamma_p \le 20$.  This indicates the model which was originally derived for ellipsoids can be approximately applied to other shapes, in particular cylinders. This fit yields $D_g=0.4 \gamma_g^{-1.2}$, consistent with values obtained for single-piece solids over a smaller parameter range \cite{chen_fluxmetric_2006}, and $D_p=0.16 \gamma_p^{-4.4}$ to characterize the particle aspect ratio dependence for the first time, to our knowledge.  However, for non-spherical particles randomly oriented in suspensions, the model prediction overestimates the measurements by a factor of 4 for $\gamma_g=4.1$.  As a result of these lower values of $\chi_{eff}$ for randomly aligned cylinders, we find $\chi_{eff}$ to be only 20\% higher for aspect ratio $\gamma_p=6.3$ cylinders than spheres.  This effect from particle misalignment remains to be included in models.  The largest  effective susceptibility we found was $\chi_{eff}=4.8$ for cylinders of particle aspect ratio $\gamma_p=6.3$ and sample aspect ratio $\gamma_g=4.1$.  We also observed that $\chi_{eff}(\gamma_p)$ displays a local maximum at $\gamma_p \approx 5$ for $\gamma_g=4.1$.  This feature is unexpected, as it was predicted that the maximum $\chi_{eff}$ would increase monotonically with $\gamma_p$ \cite{skomski_effective_2007}.   Another prediction that $\chi_{eff}$ would diverge to approach the material susceptibility $\chi$ at the liquid-solid transition $\phi_c$ \cite{martin_magnetic_2000} fails dramatically, as we observe only $\chi_{eff} \approx 4$ in the limit of this transition.  

%applications
This failure to achieve the predicted $\chi_{eff}$ approaching $\chi$ for large packing fraction and/or large aspect ratio may limit applications of magnetic suspensions,  as  $\chi_{eff} \approx \chi$  would have allowed for much stronger magnetic responses of suspensions, comparable to ferromagnetic materials.   Nonetheless, we do find  a significant range of tunable magnetic properties of magnetic suspensions up to $\chi_{eff}\approx 4$ for spherical particles,   several orders-of-magnitude stronger than other paramagnetic fluids, which typically have $\chi$ in the range of $10^{-9}$ to $10^{-4}$.  The linearity of the magnetic response without hysteresis, like paramagnetic materials, can also be desirable for simple control.

\section{Acknowledgments}      
    
    We thank Ethan Kyzivat for helping out with preliminary experiments, and ackowledge financial support from NSF Grant No. CBET-1255541 and AFOSR FA 9550-14-1-0337.

%
%\begin{equation}    
%\epsilon_{noise} \approx 0
%\end{equation}
%
%\begin{equation}
%\chi_{eff} \approx \frac{\epsilon_{ind}}{\omega I_p M_1 \alpha}
%\end{equation}
%
%\begin{equation}
%\Delta\chi_{eff} \approx \frac{\Delta\epsilon_{ind}}{\omega I_p M_1 \alpha}
%\end{equation}
%
%\begin{equation}
%\Delta\chi_{eff} \approx \frac{\Delta\epsilon_{ind}}{10^3 \times 1.25 \times 10^{-4} \times 1 \times 65 \times 10^{-3}}
%\end{equation}
%
%\begin{equation}
%\Delta \epsilon = 0.06\%Reading + 0.04\%Range
%\end{equation}
%
%\begin{equation}
%\Delta \epsilon = 0.1 mV
%\end{equation}
%
%\begin{equation}
%\Delta I = 0.10\%Reading + 0.04\%Range
%\end{equation}
%
%\begin{equation}
%\Delta I = 0.465 mA
%\end{equation}
%
%\begin{equation}
%\Delta\chi_{eff} \approx \frac{\epsilon_{ind}}{\omega M_1 \alpha} \frac{I^2}{\Delta I}
%\end{equation}
%
%\begin{equation}
%\Delta\chi_{eff} \approx \frac{100 \times 10^{-3}}{10^3 \times 1.25 \times 10^{-4} \times 1} \frac{0.465 \times 10^{-3}}{(65 \times 10^{-3})^2}
%\end{equation}
%
%
%%

%\bibliographystyle{apsrev4-1} % Tell bibtex which bibliography style to use
%\bibliography{ref_eff_sus} % Tell bibtex which .bib file to use (this one is some example file in TexLive's file tree

\begin{thebibliography}{20}%
\makeatletter
\providecommand \@ifxundefined [1]{%
 \@ifx{#1\undefined}
}%
\providecommand \@ifnum [1]{%
 \ifnum #1\expandafter \@firstoftwo
 \else \expandafter \@secondoftwo
 \fi
}%
\providecommand \@ifx [1]{%
 \ifx #1\expandafter \@firstoftwo
 \else \expandafter \@secondoftwo
 \fi
}%
\providecommand \natexlab [1]{#1}%
\providecommand \enquote  [1]{``#1''}%
\providecommand \bibnamefont  [1]{#1}%
\providecommand \bibfnamefont [1]{#1}%
\providecommand \citenamefont [1]{#1}%
\providecommand \href@noop [0]{\@secondoftwo}%
\providecommand \href [0]{\begingroup \@sanitize@url \@href}%
\providecommand \@href[1]{\@@startlink{#1}\@@href}%
\providecommand \@@href[1]{\endgroup#1\@@endlink}%
\providecommand \@sanitize@url [0]{\catcode `\\12\catcode `\$12\catcode
  `\&12\catcode `\#12\catcode `\^12\catcode `\_12\catcode `\%12\relax}%
\providecommand \@@startlink[1]{}%
\providecommand \@@endlink[0]{}%
\providecommand \url  [0]{\begingroup\@sanitize@url \@url }%
\providecommand \@url [1]{\endgroup\@href {#1}{\urlprefix }}%
\providecommand \urlprefix  [0]{URL }%
\providecommand \Eprint [0]{\href }%
\providecommand \doibase [0]{http://dx.doi.org/}%
\providecommand \selectlanguage [0]{\@gobble}%
\providecommand \bibinfo  [0]{\@secondoftwo}%
\providecommand \bibfield  [0]{\@secondoftwo}%
\providecommand \translation [1]{[#1]}%
\providecommand \BibitemOpen [0]{}%
\providecommand \bibitemStop [0]{}%
\providecommand \bibitemNoStop [0]{.\EOS\space}%
\providecommand \EOS [0]{\spacefactor3000\relax}%
\providecommand \BibitemShut  [1]{\csname bibitem#1\endcsname}%
\let\auto@bib@innerbib\@empty
%</preamble>
\bibitem [{\citenamefont {Skomski}\ \emph {et~al.}(2007)\citenamefont
  {Skomski}, \citenamefont {Hadjipanayis},\ and\ \citenamefont
  {Sellmyer}}]{skomski_effective_2007}%
  \BibitemOpen
  \bibfield  {author} {\bibinfo {author} {\bibfnamefont {R.}~\bibnamefont
  {Skomski}}, \bibinfo {author} {\bibfnamefont {G.~C.}\ \bibnamefont
  {Hadjipanayis}}, \ and\ \bibinfo {author} {\bibfnamefont {D.~J.}\
  \bibnamefont {Sellmyer}},\ }\href
  {http://cat.inist.fr/?aModele=afficheN&cpsidt=18827087} {\bibfield  {journal}
  {\bibinfo  {journal} {{IEEE} transactions on magnetics}\ }\textbf {\bibinfo
  {volume} {43}},\ \bibinfo {pages} {2956} (\bibinfo {year}
  {2007})}\BibitemShut {NoStop}%
\bibitem [{\citenamefont
  {Rosensweig}(2013)}]{rosensweig_ferrohydrodynamics_2013}%
  \BibitemOpen
  \bibfield  {author} {\bibinfo {author} {\bibfnamefont {R.~E.}\ \bibnamefont
  {Rosensweig}},\ }\href@noop {} {\emph {\bibinfo {title}
  {Ferrohydrodynamics}}}\ (\bibinfo  {publisher} {Courier Corporation},\
  \bibinfo {year} {2013})\BibitemShut {NoStop}%
\bibitem [{\citenamefont {de~Vicente}\ \emph {et~al.}(2011)\citenamefont
  {de~Vicente}, \citenamefont {Klingenberg},\ and\ \citenamefont
  {Hidalgo-Alvarez}}]{vicente_magnetorheological_2011}%
  \BibitemOpen
  \bibfield  {author} {\bibinfo {author} {\bibfnamefont {J.}~\bibnamefont
  {de~Vicente}}, \bibinfo {author} {\bibfnamefont {D.~J.}\ \bibnamefont
  {Klingenberg}}, \ and\ \bibinfo {author} {\bibfnamefont {R.}~\bibnamefont
  {Hidalgo-Alvarez}},\ }\href {\doibase 10.1039/C0SM01221A} {\bibfield
  {journal} {\bibinfo  {journal} {Soft Matter}\ }\textbf {\bibinfo {volume}
  {7}},\ \bibinfo {pages} {3701} (\bibinfo {year} {2011})}\BibitemShut
  {NoStop}%
\bibitem [{\citenamefont {Monchaux}\ \emph {et~al.}(2007)\citenamefont
  {Monchaux}, \citenamefont {Berhanu}, \citenamefont {Bourgoin}, \citenamefont
  {Moulin}, \citenamefont {Odier}, \citenamefont {Pinton}, \citenamefont
  {Volk}, \citenamefont {Fauve}, \citenamefont {Mordant}, \citenamefont
  {P{\'e}tr{\'e}lis}, \citenamefont {Chiffaudel}, \citenamefont {Daviaud},
  \citenamefont {Dubrulle}, \citenamefont {Gasquet}, \citenamefont
  {Mari{\'e}},\ and\ \citenamefont {Ravelet}}]{monchaux07}%
  \BibitemOpen
  \bibfield  {author} {\bibinfo {author} {\bibfnamefont {R.}~\bibnamefont
  {Monchaux}}, \bibinfo {author} {\bibfnamefont {M.}~\bibnamefont {Berhanu}},
  \bibinfo {author} {\bibfnamefont {M.}~\bibnamefont {Bourgoin}}, \bibinfo
  {author} {\bibfnamefont {M.}~\bibnamefont {Moulin}}, \bibinfo {author}
  {\bibfnamefont {P.}~\bibnamefont {Odier}}, \bibinfo {author} {\bibfnamefont
  {J.-F.}\ \bibnamefont {Pinton}}, \bibinfo {author} {\bibfnamefont
  {R.}~\bibnamefont {Volk}}, \bibinfo {author} {\bibfnamefont {S.}~\bibnamefont
  {Fauve}}, \bibinfo {author} {\bibfnamefont {N.}~\bibnamefont {Mordant}},
  \bibinfo {author} {\bibfnamefont {F.}~\bibnamefont {P{\'e}tr{\'e}lis}},
  \bibinfo {author} {\bibfnamefont {A.}~\bibnamefont {Chiffaudel}}, \bibinfo
  {author} {\bibfnamefont {F.}~\bibnamefont {Daviaud}}, \bibinfo {author}
  {\bibfnamefont {B.}~\bibnamefont {Dubrulle}}, \bibinfo {author}
  {\bibfnamefont {C.}~\bibnamefont {Gasquet}}, \bibinfo {author} {\bibfnamefont
  {L.}~\bibnamefont {Mari{\'e}}}, \ and\ \bibinfo {author} {\bibfnamefont
  {F.}~\bibnamefont {Ravelet}},\ }\href {\doibase
  10.1103/PhysRevLett.98.044502} {\bibfield  {journal} {\bibinfo  {journal}
  {Physical Review Letters}\ }\textbf {\bibinfo {volume} {98}},\ \bibinfo
  {pages} {044502} (\bibinfo {year} {2007})}\BibitemShut {NoStop}%
\bibitem [{\citenamefont {Stieglitz}\ and\ \citenamefont
  {M{\"u}ller}(2001)}]{stieglitz01}%
  \BibitemOpen
  \bibfield  {author} {\bibinfo {author} {\bibfnamefont {R.}~\bibnamefont
  {Stieglitz}}\ and\ \bibinfo {author} {\bibfnamefont {U.}~\bibnamefont
  {M{\"u}ller}},\ }\href {\doibase 10.1063/1.1331315} {\bibfield  {journal}
  {\bibinfo  {journal} {Physics of Fluids}\ }\textbf {\bibinfo {volume} {13}},\
  \bibinfo {pages} {561} (\bibinfo {year} {2001})}\BibitemShut {NoStop}%
\bibitem [{\citenamefont {Gailitis}\ \emph {et~al.}(2000)\citenamefont
  {Gailitis}, \citenamefont {Lielausis}, \citenamefont {Dement'ev},
  \citenamefont {Platacis}, \citenamefont {Cifersons}, \citenamefont {Gerbeth},
  \citenamefont {Gundrum}, \citenamefont {Stefani}, \citenamefont {Christen},
  \citenamefont {Hanel},\ and\ \citenamefont {Will}}]{gailitis00}%
  \BibitemOpen
  \bibfield  {author} {\bibinfo {author} {\bibfnamefont {A.}~\bibnamefont
  {Gailitis}}, \bibinfo {author} {\bibfnamefont {O.}~\bibnamefont {Lielausis}},
  \bibinfo {author} {\bibfnamefont {S.}~\bibnamefont {Dement'ev}}, \bibinfo
  {author} {\bibfnamefont {E.}~\bibnamefont {Platacis}}, \bibinfo {author}
  {\bibfnamefont {A.}~\bibnamefont {Cifersons}}, \bibinfo {author}
  {\bibfnamefont {G.}~\bibnamefont {Gerbeth}}, \bibinfo {author} {\bibfnamefont
  {T.}~\bibnamefont {Gundrum}}, \bibinfo {author} {\bibfnamefont
  {F.}~\bibnamefont {Stefani}}, \bibinfo {author} {\bibfnamefont
  {M.}~\bibnamefont {Christen}}, \bibinfo {author} {\bibfnamefont
  {H.}~\bibnamefont {Hanel}}, \ and\ \bibinfo {author} {\bibfnamefont
  {G.}~\bibnamefont {Will}},\ }\href {\doibase 10.1103/PhysRevLett.84.4365}
  {\bibfield  {journal} {\bibinfo  {journal} {Physical Review Letters}\
  }\textbf {\bibinfo {volume} {84}},\ \bibinfo {pages} {4365} (\bibinfo {year}
  {2000})}\BibitemShut {NoStop}%
\bibitem [{\citenamefont {Carle}\ \emph {et~al.}(2017)\citenamefont {Carle},
  \citenamefont {Bai}, \citenamefont {Casara}, \citenamefont {Vanderlick},\
  and\ \citenamefont {Brown}}]{CBCVB17}%
  \BibitemOpen
  \bibfield  {author} {\bibinfo {author} {\bibfnamefont {F.}~\bibnamefont
  {Carle}}, \bibinfo {author} {\bibfnamefont {K.}~\bibnamefont {Bai}}, \bibinfo
  {author} {\bibfnamefont {J.}~\bibnamefont {Casara}}, \bibinfo {author}
  {\bibfnamefont {K.}~\bibnamefont {Vanderlick}}, \ and\ \bibinfo {author}
  {\bibfnamefont {E.}~\bibnamefont {Brown}},\ }\href@noop {} {\bibfield
  {journal} {\bibinfo  {journal} {Phys. Rev. Fluids}\ }\textbf {\bibinfo
  {volume} {2}},\ \bibinfo {pages} {013301} (\bibinfo {year}
  {2017})}\BibitemShut {NoStop}%
\bibitem [{\citenamefont {Aharoni}(1998)}]{aharoni_demagnetizing_1998}%
  \BibitemOpen
  \bibfield  {author} {\bibinfo {author} {\bibfnamefont {A.}~\bibnamefont
  {Aharoni}},\ }\href {\doibase 10.1063/1.367113} {\bibfield  {journal}
  {\bibinfo  {journal} {Journal of Applied Physics}\ }\textbf {\bibinfo
  {volume} {83}},\ \bibinfo {pages} {3432} (\bibinfo {year}
  {1998})}\BibitemShut {NoStop}%
\bibitem [{\citenamefont {de~Vicente}\ \emph {et~al.}(2010)\citenamefont
  {de~Vicente}, \citenamefont {Vereda}, \citenamefont {Segovia-Guti{\'e}rrez},
  \citenamefont {Morales},\ and\ \citenamefont
  {Hidalgo-Álvarez}}]{vicente_effect_2010}%
  \BibitemOpen
  \bibfield  {author} {\bibinfo {author} {\bibfnamefont {J.}~\bibnamefont
  {de~Vicente}}, \bibinfo {author} {\bibfnamefont {F.}~\bibnamefont {Vereda}},
  \bibinfo {author} {\bibfnamefont {J.~P.}\ \bibnamefont
  {Segovia-Guti{\'e}rrez}}, \bibinfo {author} {\bibfnamefont {M.~d.~P.}\
  \bibnamefont {Morales}}, \ and\ \bibinfo {author} {\bibfnamefont
  {R.}~\bibnamefont {Hidalgo-Álvarez}},\ }\href {\doibase 10.1122/1.3479045}
  {\bibfield  {journal} {\bibinfo  {journal} {Journal of Rheology
  (1978-present)}\ }\textbf {\bibinfo {volume} {54}},\ \bibinfo {pages} {1337}
  (\bibinfo {year} {2010})}\BibitemShut {NoStop}%
\bibitem [{\citenamefont {Bleaney}\ and\ \citenamefont
  {Hull}(1941)}]{bleaney_effective_1941}%
  \BibitemOpen
  \bibfield  {author} {\bibinfo {author} {\bibfnamefont {B.}~\bibnamefont
  {Bleaney}}\ and\ \bibinfo {author} {\bibfnamefont {R.~A.}\ \bibnamefont
  {Hull}},\ }\href {\doibase 10.1098/rspa.1941.0045} {\bibfield  {journal}
  {\bibinfo  {journal} {Proceedings of the Royal Society of London A:
  Mathematical, Physical and Engineering Sciences}\ }\textbf {\bibinfo {volume}
  {178}},\ \bibinfo {pages} {86} (\bibinfo {year} {1941})}\BibitemShut
  {NoStop}%
\bibitem [{\citenamefont {Bj{\o}rk}\ and\ \citenamefont
  {Bahl}(2013)}]{bjork_demagnetization_2013}%
  \BibitemOpen
  \bibfield  {author} {\bibinfo {author} {\bibfnamefont {R.}~\bibnamefont
  {Bj{\o}rk}}\ and\ \bibinfo {author} {\bibfnamefont {C.~R.~H.}\ \bibnamefont
  {Bahl}},\ }\href {\doibase 10.1063/1.4820141} {\bibfield  {journal} {\bibinfo
   {journal} {Applied Physics Letters}\ }\textbf {\bibinfo {volume} {103}},\
  \bibinfo {pages} {102403} (\bibinfo {year} {2013})}\BibitemShut {NoStop}%
\bibitem [{\citenamefont {Martin}\ \emph {et~al.}(2000)\citenamefont {Martin},
  \citenamefont {Odier}, \citenamefont {Pinton},\ and\ \citenamefont
  {Fauve}}]{martin_magnetic_2000}%
  \BibitemOpen
  \bibfield  {author} {\bibinfo {author} {\bibfnamefont {A.}~\bibnamefont
  {Martin}}, \bibinfo {author} {\bibfnamefont {P.}~\bibnamefont {Odier}},
  \bibinfo {author} {\bibfnamefont {J.-F.}\ \bibnamefont {Pinton}}, \ and\
  \bibinfo {author} {\bibfnamefont {S.}~\bibnamefont {Fauve}},\ }\href
  {\doibase 10.1007/s100510070066} {\bibfield  {journal} {\bibinfo  {journal}
  {The European Physical Journal B - Condensed Matter and Complex Systems}\
  }\textbf {\bibinfo {volume} {18}},\ \bibinfo {pages} {337} (\bibinfo {year}
  {2000})}\BibitemShut {NoStop}%
\bibitem [{\citenamefont {Frick}\ \emph {et~al.}(2002)\citenamefont {Frick},
  \citenamefont {Khripchenko}, \citenamefont {Denisov}, \citenamefont
  {Sokoloff},\ and\ \citenamefont {Pinton}}]{frick_effective_2002}%
  \BibitemOpen
  \bibfield  {author} {\bibinfo {author} {\bibfnamefont {P.}~\bibnamefont
  {Frick}}, \bibinfo {author} {\bibfnamefont {S.}~\bibnamefont {Khripchenko}},
  \bibinfo {author} {\bibfnamefont {S.}~\bibnamefont {Denisov}}, \bibinfo
  {author} {\bibfnamefont {D.}~\bibnamefont {Sokoloff}}, \ and\ \bibinfo
  {author} {\bibfnamefont {J.-F.}\ \bibnamefont {Pinton}},\ }\href {\doibase
  10.1140/epjb/e20020044} {\bibfield  {journal} {\bibinfo  {journal} {The
  European Physical Journal B - Condensed Matter and Complex Systems}\ }\textbf
  {\bibinfo {volume} {25}},\ \bibinfo {pages} {399} (\bibinfo {year}
  {2002})}\BibitemShut {NoStop}%
\bibitem [{\citenamefont {Dickey}\ \emph {et~al.}(2008)\citenamefont {Dickey},
  \citenamefont {Chiechi}, \citenamefont {Larsen}, \citenamefont {Weiss},
  \citenamefont {Weitz},\ and\ \citenamefont {Whitesides}}]{DCLWWW08}%
  \BibitemOpen
  \bibfield  {author} {\bibinfo {author} {\bibfnamefont {M.~D.}\ \bibnamefont
  {Dickey}}, \bibinfo {author} {\bibfnamefont {R.~C.}\ \bibnamefont {Chiechi}},
  \bibinfo {author} {\bibfnamefont {R.~J.}\ \bibnamefont {Larsen}}, \bibinfo
  {author} {\bibfnamefont {E.~A.}\ \bibnamefont {Weiss}}, \bibinfo {author}
  {\bibfnamefont {D.~A.}\ \bibnamefont {Weitz}}, \ and\ \bibinfo {author}
  {\bibfnamefont {G.~M.}\ \bibnamefont {Whitesides}},\ }\href {\doibase
  10.1002/adfm.200701216} {\bibfield  {journal} {\bibinfo  {journal} {Adv.
  Funct. Mater.}\ }\textbf {\bibinfo {volume} {18}},\ \bibinfo {pages} {1097}
  (\bibinfo {year} {2008})}\BibitemShut {NoStop}%
\bibitem [{\citenamefont {Chen}\ \emph {et~al.}(2006)\citenamefont {Chen},
  \citenamefont {Pardo},\ and\ \citenamefont {Sanchez}}]{chen_fluxmetric_2006}%
  \BibitemOpen
  \bibfield  {author} {\bibinfo {author} {\bibfnamefont {D.~X.}\ \bibnamefont
  {Chen}}, \bibinfo {author} {\bibfnamefont {E.}~\bibnamefont {Pardo}}, \ and\
  \bibinfo {author} {\bibfnamefont {A.}~\bibnamefont {Sanchez}},\ }\href
  {\doibase 10.1016/j.jmmm.2006.02.235} {\bibfield  {journal} {\bibinfo
  {journal} {Journal of Magnetism and Magnetic Materials}\ }\textbf {\bibinfo
  {volume} {306}},\ \bibinfo {pages} {135} (\bibinfo {year}
  {2006})}\BibitemShut {NoStop}%
\bibitem [{\citenamefont {Desmond}\ and\ \citenamefont {Weeks}(2009)}]{DW09}%
  \BibitemOpen
  \bibfield  {author} {\bibinfo {author} {\bibfnamefont {K.~W.}\ \bibnamefont
  {Desmond}}\ and\ \bibinfo {author} {\bibfnamefont {E.~R.}\ \bibnamefont
  {Weeks}},\ }\href@noop {} {\bibfield  {journal} {\bibinfo  {journal} {Phys.
  Rev. E}\ }\textbf {\bibinfo {volume} {80}},\ \bibinfo {pages} {051305}
  (\bibinfo {year} {2009})}\BibitemShut {NoStop}%
\bibitem [{\citenamefont {Brown}\ \emph {et~al.}(2010)\citenamefont {Brown},
  \citenamefont {Zhang}, \citenamefont {Forman}, \citenamefont {Maynor},
  \citenamefont {Betts}, \citenamefont {DeSimone},\ and\ \citenamefont
  {Jaeger}}]{BZFMBDJ10}%
  \BibitemOpen
  \bibfield  {author} {\bibinfo {author} {\bibfnamefont {E.}~\bibnamefont
  {Brown}}, \bibinfo {author} {\bibfnamefont {H.}~\bibnamefont {Zhang}},
  \bibinfo {author} {\bibfnamefont {N.~A.}\ \bibnamefont {Forman}}, \bibinfo
  {author} {\bibfnamefont {B.~W.}\ \bibnamefont {Maynor}}, \bibinfo {author}
  {\bibfnamefont {D.~E.}\ \bibnamefont {Betts}}, \bibinfo {author}
  {\bibfnamefont {J.~M.}\ \bibnamefont {DeSimone}}, \ and\ \bibinfo {author}
  {\bibfnamefont {H.~M.}\ \bibnamefont {Jaeger}},\ }\href@noop {} {\bibfield
  {journal} {\bibinfo  {journal} {J. Rheology}\ }\textbf {\bibinfo {volume}
  {54}},\ \bibinfo {pages} {1023} (\bibinfo {year} {2010})}\BibitemShut
  {NoStop}%
\bibitem [{\citenamefont {Brown}\ \emph {et~al.}(2011)\citenamefont {Brown},
  \citenamefont {Zhang}, \citenamefont {Forman}, \citenamefont {Maynor},
  \citenamefont {Betts}, \citenamefont {DeSimone},\ and\ \citenamefont
  {Jaeger}}]{brown_shear_2011}%
  \BibitemOpen
  \bibfield  {author} {\bibinfo {author} {\bibfnamefont {E.}~\bibnamefont
  {Brown}}, \bibinfo {author} {\bibfnamefont {H.}~\bibnamefont {Zhang}},
  \bibinfo {author} {\bibfnamefont {N.~A.}\ \bibnamefont {Forman}}, \bibinfo
  {author} {\bibfnamefont {B.~W.}\ \bibnamefont {Maynor}}, \bibinfo {author}
  {\bibfnamefont {D.~E.}\ \bibnamefont {Betts}}, \bibinfo {author}
  {\bibfnamefont {J.~M.}\ \bibnamefont {DeSimone}}, \ and\ \bibinfo {author}
  {\bibfnamefont {H.~M.}\ \bibnamefont {Jaeger}},\ }\href {\doibase
  10.1103/PhysRevE.84.031408} {\bibfield  {journal} {\bibinfo  {journal}
  {Physical Review E}\ }\textbf {\bibinfo {volume} {84}},\ \bibinfo {pages}
  {031408} (\bibinfo {year} {2011})}\BibitemShut {NoStop}%
\bibitem [{\citenamefont
  {Philipse}(1996)}]{Philipse_Random_Contact_Equation_96}%
  \BibitemOpen
  \bibfield  {author} {\bibinfo {author} {\bibfnamefont {A.~P.}\ \bibnamefont
  {Philipse}},\ }\href {\doibase 10.1021/la950671o} {\bibfield  {journal}
  {\bibinfo  {journal} {Langmuir}\ }\textbf {\bibinfo {volume} {12}},\ \bibinfo
  {pages} {1127–} (\bibinfo {year} {1996})}\BibitemShut {NoStop}%
\bibitem [{\citenamefont {Trappe}\ \emph {et~al.}(2001)\citenamefont {Trappe},
  \citenamefont {Prasad}, \citenamefont {Cipelletti}, \citenamefont {Segre},\
  and\ \citenamefont {Weitz}}]{trappe_jamming_2001}%
  \BibitemOpen
  \bibfield  {author} {\bibinfo {author} {\bibfnamefont {V.}~\bibnamefont
  {Trappe}}, \bibinfo {author} {\bibfnamefont {V.}~\bibnamefont {Prasad}},
  \bibinfo {author} {\bibfnamefont {L.}~\bibnamefont {Cipelletti}}, \bibinfo
  {author} {\bibfnamefont {P.~N.}\ \bibnamefont {Segre}}, \ and\ \bibinfo
  {author} {\bibfnamefont {D.~A.}\ \bibnamefont {Weitz}},\ }\href {\doibase
  10.1038/35081021} {\bibfield  {journal} {\bibinfo  {journal} {Nature}\
  }\textbf {\bibinfo {volume} {411}},\ \bibinfo {pages} {772} (\bibinfo {year}
  {2001})}\BibitemShut {NoStop}%
\end{thebibliography}


\end{document}